\pdfoutput=1
\documentclass[final,5p,times,twocolumn,nopreprintline]{elsarticle}
\usepackage{amsmath,slashed}
\usepackage{graphicx,graphics}
\usepackage{dcolumn}
\usepackage[hyperfootnotes=false]{hyperref}
\usepackage{xspace}
\usepackage{color}
\usepackage{balance}

\usepackage{fancyhdr}
\fancypagestyle{firstpage}{%
	
	\lhead{}
	\rhead{\small UWThPh 2020-25}
}
\newcommand{\GeV}{\,\text{GeV}}

\newcommand{\fm}{\,\text{fm}}

\newcommand{\EL}{E\L{}}
\newcommand{\mpi}{M_\pi}
\newcommand{\mw}{M_\omega}
\renewcommand{\Im}{\text{Im}\,}

\newcommand{\beq}{\begin{equation}}
\newcommand{\eeq}{\end{equation}}
\newcommand{\<}{\langle}
\renewcommand{\>}{\rangle}

\newcommand{\Dalphahadpipi}{\Delta\alpha_{\pi\pi}^{(5)}(M_Z^2)}
\newcommand{\amupipi}{a_\mu^{\pi\pi}\big|_{\le1\GeV}}
\newcommand{\nn}{\nonumber\\}

\def\Xint#1{\mathchoice
   {\XXint\displaystyle\textstyle{#1}}%
   {\XXint\textstyle\scriptstyle{#1}}%
   {\XXint\scriptstyle\scriptscriptstyle{#1}}%
   {\XXint\scriptscriptstyle\scriptscriptstyle{#1}}%
   \!\int}
\def\XXint#1#2#3{{\setbox0=\hbox{$#1{#2#3}{\int}$}
     \vcenter{\hbox{$#2#3$}}\kern-0.5\wd0}}
\providecommand{\dashint}[1][0pt]{\Xint{\hspace{#1}-}}

\allowdisplaybreaks[1]

\begin{document}

\renewcommand{\theequation}{\arabic{equation}}

\begin{frontmatter}

\title{Constraints on the two-pion contribution to hadronic vacuum polarization} 

\author[Bern]{Gilberto Colangelo}
\author[Bern]{Martin Hoferichter}
\author[UCSD,Vienna]{Peter Stoffer}

\address[Bern]{Albert Einstein Center for Fundamental Physics, Institute for Theoretical Physics, University of Bern, Sidlerstrasse 5, 3012 Bern, Switzerland}
\address[UCSD]{Department of Physics, University of California at San Diego, La Jolla, CA 92093, USA}
\address[Vienna]{University of Vienna, Faculty of Physics, Boltzmanngasse 5, 1090 Vienna, Austria}

\begin{abstract}
At low energies hadronic vacuum polarization (HVP) is strongly dominated by
two-pion intermediate states, which are responsible for about $70\%$ of the
HVP contribution to the anomalous magnetic moment of the muon,
$a_\mu^\text{HVP}$. Lattice-QCD evaluations of the latter indicate that it
might be larger than calculated dispersively on the basis of $e^+e^-
\to$~hadrons data, at a level which would contest the long-standing discrepancy with the
$a_\mu$ measurement. In this Letter we study to which extent
this $2\pi$ contribution can be modified without, at the same time,
producing a conflict elsewhere in low-energy hadron phenomenology. To this
end we consider a dispersive representation of the $e^+e^- \to 2\pi$
process and study the correlations which thereby emerge between
$a_\mu^\text{HVP}$, the hadronic running of the fine-structure constant,
the $P$-wave $\pi \pi$  phase shift, and the charge radius of the
pion. Inelastic effects play an important role, despite being constrained
by the Eidelman--\L{}ukaszuk bound. We identify scenarios in
which $a_\mu^\text{HVP}$ can be altered substantially, driven by changes in the phase shift and/or the inelastic contribution, 
and illustrate the
ensuing changes in the $e^+e^-\to 2\pi$ cross section. In the combined scenario, which minimizes the effect in the cross section, a uniform shift around $4\%$ is required. At the same time both the analytic continuation into the space-like region and the pion charge radius are affected at a level that could be probed in future lattice-QCD calculations.  
\end{abstract}

\end{frontmatter}

\thispagestyle{firstpage}

\section{Introduction}

The uncertainty in the Standard Model prediction for the anomalous magnetic moment of the muon~\cite{Aoyama:2020ynm,Aoyama:2012wk,Aoyama:2019ryr,Czarnecki:2002nt,Gnendiger:2013pva,Davier:2017zfy,Keshavarzi:2018mgv,Colangelo:2018mtw,Hoferichter:2019gzf,Davier:2019can,Keshavarzi:2019abf,Hoid:2020xjs,Kurz:2014wya,Melnikov:2003xd,Colangelo:2014dfa,Colangelo:2014pva,Colangelo:2015ama,Masjuan:2017tvw,Colangelo:2017qdm,Colangelo:2017fiz,Hoferichter:2018dmo,Hoferichter:2018kwz,Gerardin:2019vio,Bijnens:2019ghy,Colangelo:2019lpu,Colangelo:2019uex,Blum:2019ugy,Colangelo:2014qya}
\beq
\label{amuSM}
a_\mu^\text{SM}=116\,591\,810(43)\times 10^{-11}
\eeq
is currently dominated by HVP, whose leading-order contribution as derived from $e^+e^-\to\text{hadrons}$ cross sections reads~\cite{Aoyama:2020ynm,Davier:2017zfy,Keshavarzi:2018mgv,Colangelo:2018mtw,Hoferichter:2019gzf,Davier:2019can,Keshavarzi:2019abf,Hoid:2020xjs}
\beq
\label{ee}
a_\mu^\text{HVP}\big|_{e^+e^-}=6\,931(40)\times 10^{-11}.  
\eeq
The resulting SM prediction~\eqref{amuSM} differs from experiment~\cite{bennett:2006fi}
\beq
\label{exp}
a_\mu^\text{exp}=116\,592\,089(63)\times 10^{-11}
\eeq
by $3.7\sigma$. If this discrepancy were all to be blamed on an incorrect
evaluation of the HVP contribution, this would have to be as large as
$7\,200 \times 10^{-11}$ to reconcile the central values of the SM and
experiment. That such a possibility should in fact be seriously considered
has become a pressing issue in view of recent lattice-QCD evaluations. The
lattice average performed in Ref.~\cite{Aoyama:2020ynm} (based on
Refs.~\cite{Chakraborty:2017tqp,Borsanyi:2017zdw,Blum:2018mom,Giusti:2019xct,Shintani:2019wai,Davies:2019efs,Gerardin:2019rua,Aubin:2019usy,Giusti:2019hkz})
\beq
\label{eq:LatticeAverage}
a_\mu^\text{HVP}\big|_\text{lattice\,average}=7\,116(184)\times 10^{-11}
\eeq
is consistent with both the $e^+e^-$ value~\eqref{ee} (within $1\sigma$),
but also with the experimental value~\eqref{exp}. The more recent
calculation of Ref.~\cite{Borsanyi:2020mff},
$a_\mu^\text{HVP}=7\,087(53)\times 10^{-11}$, quotes a slightly smaller
central value, but due to the increased precision lies above the $e^+e^-$
value by $2.3\sigma$, while reducing the tension with Eq.~\eqref{exp} to
$1.5\sigma$.

For the second-most-important class of hadronic contributions, hadronic
light-by-light scattering (HLbL), the phenomenological estimate
$a_\mu^\text{HLbL}=92(19)\times
10^{-11}$~\cite{Aoyama:2020ynm,Melnikov:2003xd,Colangelo:2014dfa,Colangelo:2014pva,Colangelo:2015ama,Masjuan:2017tvw,Colangelo:2017qdm,Colangelo:2017fiz,Hoferichter:2018dmo,Hoferichter:2018kwz,Gerardin:2019vio,Bijnens:2019ghy,Colangelo:2019lpu,Colangelo:2019uex,Pauk:2014rta,Danilkin:2016hnh,Jegerlehner:2017gek,Knecht:2018sci,Eichmann:2019bqf,Roig:2019reh}
agrees with $a_\mu^\text{HLbL}=82(35)\times 10^{-11}$ from lattice
QCD~\cite{Blum:2019ugy} (including the phenomenological estimate for the
charm contribution), in such a way that an average of the two has been used
in Eq.~\eqref{amuSM}.

This situation has triggered renewed interest in the consequences of large
changes to HVP elsewhere, especially for global electroweak fits due to its
impact on the hadronic running of the fine-structure constant
$\alpha$~\cite{Passera:2008jk,Crivellin:2020zul,Keshavarzi:2020bfy,Malaescu:2020zuc}.
These analyses have shown that to avoid a significant tension with
electroweak precision data, the changes to the hadronic cross sections need
to be concentrated at low energies, at least below $2\GeV$, a scenario
indeed indicated by Ref.~\cite{Borsanyi:2020mff}.

In previous
work~\cite{Crivellin:2020zul,Keshavarzi:2020bfy,Malaescu:2020zuc} changes
to the hadronic cross sections were considered as a whole, with specific
assumptions on the energy dependence. However, if the changes are
concentrated in the low-energy region, it is clear that the most relevant
absolute effect will occur in the dominant $2\pi$ channel, since the
required relative changes in the subleading channels would become
prohibitively large. In this region, the $2 \pi$ channel is essentially
elastic and dominated by the $\rho$ resonance. The relevant hadronic matrix
element, the pion vector form factor (VFF), is strongly constrained by
analyticity and unitarity, which imply that below $1\GeV$ it is essentially
determined by the $P$-wave $\pi \pi$ phase shift~\cite{Colangelo:2018mtw},
which is again constrained by analyticity, unitarity, and crossing symmetry,
taking the form of Roy 
equations~\cite{Roy:1971tc,Ananthanarayan:2000ht,GarciaMartin:2011cn,Caprini:2011ky}. The
main conclusion of the analysis in Ref.~\cite{Colangelo:2018mtw} is that the VFF
below $1\GeV$ can be described in terms of a handful of parameters, which can
all be determined by a fit to the $e^+e^- \to 2 \pi$ data. The fact that
these data, which have now reached a remarkable level of precision,
typically below $1\%$, can be well described by this highly constrained
representation, is a nontrivial test on their quality.

Within this framework it is possible to address the question which changes
become possible without violating analyticity and unitarity and without
incurring other tensions elsewhere---besides those with the $e^+e^-\to 2\pi$
cross-section data. To this end, we first of all determine what changes in
the parameters of the dispersive representation may generate the desired
change in $a_\mu^\text{HVP}$. With the same set of parameters we then
calculate the $P$-wave $\pi \pi$ phase shifts, the hadronic running of
$\alpha$, as well as the charge radius of the pion, and thereby establish
correlations among all these quantities.

Finally, we identify scenarios in which significant changes to HVP remain
possible despite these independent constraints on the pion VFF. The
comparison of the resulting predictions for the $e^+e^-\to 2\pi$ cross
section to data allows us to quantify by how much the experimental cross
sections would need to be changed to accommodate such an increase in
$a_\mu^\text{HVP}$.

\section{The pion vector form factor}

The HVP contribution to the anomalous magnetic moment of the muon, expressed in terms of the $e^+e^-\to\text{hadrons}$ cross section, reads~\cite{Bouchiat:1961lbg,Brodsky:1967sr}
\begin{align}
\label{amu_HVP}
 a_\mu^\text{HVP}&=\bigg(\frac{\alpha m_\mu}{3\pi}\bigg)^2\int_{s_\text{thr}}^\infty ds \frac{\hat K(s)}{s^2}R_\text{had}(s),\notag\\
 R_\text{had}(s)&=\frac{3s}{4\pi\alpha^2}\sigma(e^+e^-\to\text{hadrons}),
\end{align}
with a known kernel function $\hat K(s)$. With the pion VFF $F_\pi^V(s)$ defined as the matrix element of the electromagnetic current $j_\mathrm{em}^\mu$,
\beq
\label{VFF_def}
	\langle \pi^\pm(p') | j_\mathrm{em}^\mu(0) | \pi^\pm(p) \rangle =\pm (p'+p)^\mu F_\pi^V((p'-p)^2),
\eeq
the $2\pi$ contribution becomes
\beq
	\sigma(e^+e^-\to\pi^+\pi^-) = \frac{\pi \alpha^2}{3s} \sigma_\pi^3(s) \big| F_\pi^V(s) \big|^2 ,
\eeq
where $\sigma_\pi(s) = \sqrt{1-4\mpi^2/s}$. Similarly, the two-pion contribution to the hadronic running of $\alpha$, evaluated at $M_Z^2$, 
\beq
\label{had_running}
 \Delta \alpha^{(5)}_\text{had}(M_Z^2)=\frac{\alpha M_Z^2}{3\pi}\dashint[0.5pt]^\infty_{s_\text{thr}}d s \frac{R_\text{had}(s)}{s(M_Z^2-s)},
\eeq
is determined by $F_\pi^V(s)$. In both cases, the integration threshold becomes $s_\text{thr}=4\mpi^2$, and radiative corrections to the cross section are implemented in such a way that vacuum polarization is removed, but final-state radiation (FSR) included. Since Eq.~\eqref{VFF_def} defines the matrix element in pure QCD, this implies that FSR corrections need to be included in the final step, see Ref.~\cite{Colangelo:2018mtw} for further details. In addition, we consider the correlation with the pion charge radius
\beq
	\label{eq:rpiSumRule}
	\langle r_\pi^2\rangle=6\frac{d F_\pi^V(s)}{d s}\bigg|_{s=0}=\frac{6}{\pi}\int_{4\mpi^2}^\infty d s\frac{\Im F_\pi^V(s)}{s^2},
\eeq
which, contrary to $a_\mu^\text{HVP}$ and $\Delta\alpha_\text{had}^{(5)}$,
is also explicitly sensitive to the phase of $F_\pi^V(s)$. 

In the elastic region, where $2\pi$ is again the only relevant intermediate
state, $F_\pi^V(s)$ is strongly constrained by analyticity and
unitarity. If the elastic region extended all the way to infinity, the
solution to the unitarity and analyticity constraints would be given by the
Omn\`es factor~\cite{Omnes:1958hv}
\beq
\label{omnes}
\Omega_1^1(s) = \exp\left\{ \frac{s}{\pi} \int_{4\mpi^2}^\infty ds^\prime \frac{\delta_1^1(s^\prime)}{s^\prime(s^\prime-s)} \right\},
\eeq
with the $P$-wave $\pi\pi$ scattering phase shift $\delta_1^1(s)$. This
phase shift, in turn, is strongly constrained by $\pi\pi$ Roy
equations~\cite{Roy:1971tc,Ananthanarayan:2000ht,GarciaMartin:2011cn,Caprini:2011ky},
which further limits the permissible changes in $F_\pi^V(s)$, see
Refs.~\cite{DeTroconiz:2001rip,Leutwyler:2002hm,Colangelo:2003yw,deTroconiz:2004yzs,Hoferichter:2016duk,Hanhart:2016pcd,Ananthanarayan:2018nyx,Ananthanarayan:2020vum}
for representations that exploit this intimate connection between the VFF
and $\pi\pi$ scattering. Below $1\GeV$ inelastic effects are small, but at
the level of precision necessary here, have to be taken into account. To do
this we multiply the fully elastic Omn\`es factor~\eqref{omnes} by two
additional factors, as in
Refs.~\cite{Leutwyler:2002hm,Colangelo:2003yw,Colangelo:2018mtw}   
\beq
	\label{eq:PionVFF}
	F_\pi^V(s) = \Omega_1^1(s) G_\omega(s) G_\mathrm{in}^N(s),
\eeq
where $G_\omega(s)$ accounts for the isospin-violating $3\pi$ cut, which is
completely dominated by $\rho$--$\omega$ mixing, and the $4\pi$ cut is expanded
into a conformal polynomial 
\beq
\label{Gin_def}
	G_\mathrm{in}^N(s) = 1 + \sum_{k=1}^N c_k ( z^k(s) - z^k(0) ),
\eeq
where the conformal variable 
\beq
	z(s) = \frac{\sqrt{s_\mathrm{in} - s_c} - \sqrt{s_\mathrm{in} - s}}{\sqrt{s_\mathrm{in} - s_c} + \sqrt{s_\mathrm{in} - s}}
\eeq
permits inelastic phases above the $\pi\omega$ threshold $s_\mathrm{in} =
(M_{\pi^0} + \mw)^2$. The parameter $s_c$ is the value of $s$ mapped to the
origin, $z(s_c)=0$, and is varied around $-1\GeV^2$. To ensure the 
correct threshold behavior, the $c_k$ are related by an additional
constraint that removes the $S$-wave singularity.  

In total, the dispersive representation from Ref.~\cite{Colangelo:2018mtw} then involves the following free parameters: first, the solution of the $\pi\pi$ Roy equations is determined once the phase shifts at $s_0=(0.8\GeV)^2$ and $s_1=(1.15\GeV)^2$ are specified, so that $\delta^1_1(s_0)$ and $\delta^1_1(s_1)$ are free fit parameters. Second, $G_\omega(s)$ depends on the $\omega$ pole parameters as well as the overall strength of $\rho$--$\omega$ mixing. Third, there are $N-1$ free parameters in $G_\mathrm{in}^N(s)$ to describe inelastic effects.

The results for the phase shifts from a fit to VFF data are~\cite{Colangelo:2018mtw}
\beq
\label{pipi_phase_final}
\delta^1_1(s_0)=110.4(7)^\circ,\qquad 
\delta^1_1(s_1)=165.7(2.4)^\circ, 
\eeq
but for the purpose of this work it is crucial to understand to within which ranges they can be constrained without relying on $e^+e^-\to 2\pi$ data (or $\tau\to\pi\pi\nu_\tau$). In principle, one could even consider indirect constraints that arise, via the Roy equations, from low-energy data in crossed channels, such as $K_{\ell 4}$ data~\cite{Batley:2010zza,Batley:2012rf,Colangelo:2015kha}, but here we simply quote the results from the partial-wave analyses
\begin{align}
\label{PWA}
	\begin{alignedat}{3}
		&\text{Ref.~\cite{Hyams:1973zf}}:			&	\delta^1_1(0.79\GeV)&=97.5(1.5)^\circ	\quad & & [103.9(6)^\circ], \\
		&									&	\delta^1_1(0.81\GeV)&=112.1(8)^\circ	\quad & & [116.2(7)^\circ], \\
		&									&	\delta^1_1(1.15\GeV)&=167.7(3.3)^\circ	\quad & & [165.7(2.4)^\circ], \\
		&\text{Ref.~\cite{Protopopescu:1973sh}}: \quad	&	\delta^1_1(0.795\GeV)&=105.0(1.5)^\circ	\quad & & [107.2(6)^\circ], \\
		&									&	\delta^1_1(0.81\GeV)&=114.0(1.4)^\circ	\quad & & [116.2(7)^\circ], \\
		&									&	\delta^1_1(1.15\GeV)&=164(6)^\circ		\quad & & [165.7(2.4)^\circ],
	\end{alignedat}
\end{align}
where our values, extracted from the global fit to $e^+e^-\to2\pi$ data,
are shown in brackets for comparison. 

The parameters in $G_\omega(s)$ do not need to be considered further because either one would have to be changed beyond any plausible range to produce a relevant effect in $a_\mu^\text{HVP}$. Finally, if several free parameters in the conformal polynomial are introduced, the resulting inelastic phase shift in general leads to unacceptably large violations of Watson's final-state theorem~\cite{Watson:1954uc}. A quantitative phenomenological bound can be formulated based on the ratio
\beq
	r = \frac{\sigma^{I=1}(e^+e^-\to\mathrm{hadrons})}{\sigma(e^+e^-\to\pi^+\pi^-)} - 1
\eeq
of non-$2\pi$ to $2\pi$ hadronic cross sections for isospin $I=1$, e.g., for the total phase $\psi$ of the VFF~\cite{Lukaszuk:1973jd,Eidelman:2003uh}
\beq
	\label{eq:ELBoundAlt}
	\sin^2(\psi-\delta_1^1) \le \frac{1}{2} \Big( 1-\sqrt{1-r^2} \Big).
\eeq
This \EL{} bound shows that inelastic effects below the $\pi\omega$
threshold are indeed negligible, and limits the size of the inelastic phase
above. In practice, we use the implementation of the \EL{} bound from
Ref.~\cite{Colangelo:2018mtw}, but note that these details are of limited
importance in the present context: once the \EL{} bound becomes active, the
increase in the $\chi^2$ is rather steep, so that the excluded parameter
space is essentially insensitive to the exact implementation of the \EL{}
bound.

\section{Changing HVP}
\label{sec:ChangingHVP}

We start from the main results of Ref.~\cite{Colangelo:2018mtw}, where the representation~\eqref{eq:PionVFF} is fit to a combination of the data sets of Refs.~\cite{Amendolia:1986wj,Akhmetshin:2001ig,Akhmetshin:2003zn,Achasov:2005rg,Achasov:2006vp,Akhmetshin:2006wh,Akhmetshin:2006bx,Ambrosino:2008aa,Aubert:2009ad,Ambrosino:2010bv,Lees:2012cj,Babusci:2012rp,Anastasi:2017eio}, leading to a two-pion contribution to $a_\mu^\text{HVP}$ below $1\GeV$ of~\cite{Colangelo:2018mtw}
\begin{align}
	\label{eq:CentralAmu}
	\amupipi &= 495.0(1.5)(2.1) \times 10^{-10} \nn
		&= 495.0(2.6) \times 10^{-10} ,
\end{align}
where the first error is the fit uncertainty (inflated by $\sqrt{\chi^2/\mathrm{dof}}$) and the second error includes all systematic uncertainties of the representation~\eqref{eq:PionVFF}. The central configuration uses $N-1=4$ free parameters in the conformal polynomial. Due to the sensitivity of the radius sum rule~\eqref{eq:rpiSumRule} to the phase of the VFF, fits with too many free parameters in the conformal polynomial tend to become unstable for $\< r_\pi^2 \>$, because the phase needs to be extrapolated above the energy for which the \EL{} bound can be used in practice to constrain the size of the imaginary part. For this reason, in Ref.~\cite{Colangelo:2018mtw} the central evaluation of $\< r_\pi^2\>$ was obtained with $N-1=1$, but the full variation with $N$ was kept as a systematic uncertainty, which dominates the uncertainty assigned to the final result~\cite{Colangelo:2018mtw}
\beq
	\< r_\pi^2 \> = 0.429(1)(4) \fm^2 = 0.429(4) \fm^2 .
\eeq
Here, we use as reference point the value for $N-1=4$~\cite{Colangelo:2018mtw}
\beq
	\< r_\pi^2 \>\big|_{N-1=4} = 0.426(1) \fm^2,
\eeq
where the error refers to the fit uncertainty only.
Finally, the fit configuration with $N-1=4$ leads to a two-pion contribution to the hadronic running of $\alpha$, $\Dalphahadpipi$, of
\begin{align}
	\Dalphahadpipi \big|_{\le1\GeV} &= 32.62(10)(11) \times 10^{-4} \nn
		&= 32.62(15) \times 10^{-4} \, .
\end{align}

\begin{figure}[t]
	\centering
	\scalebox{0.8}{
	\input{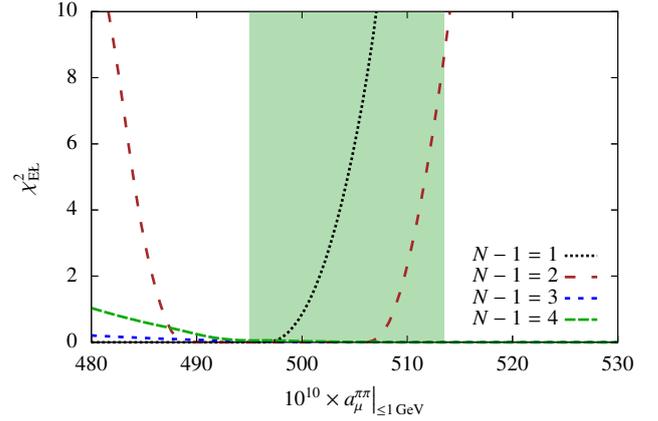}}%
	\caption{Impact of the \EL{} bound on the $\chi^2$ for $N-1=1\ldots4$ when varying $\amupipi$ away from the central fit result. The shaded area corresponds to $0\leq\Delta a_\mu^{\pi\pi} \big|_{\le1\GeV} \le 18.5 \times 10^{-10}$ for $N-1=4$.}
	\label{img:EL1}
\end{figure}

\begin{figure*}[t]
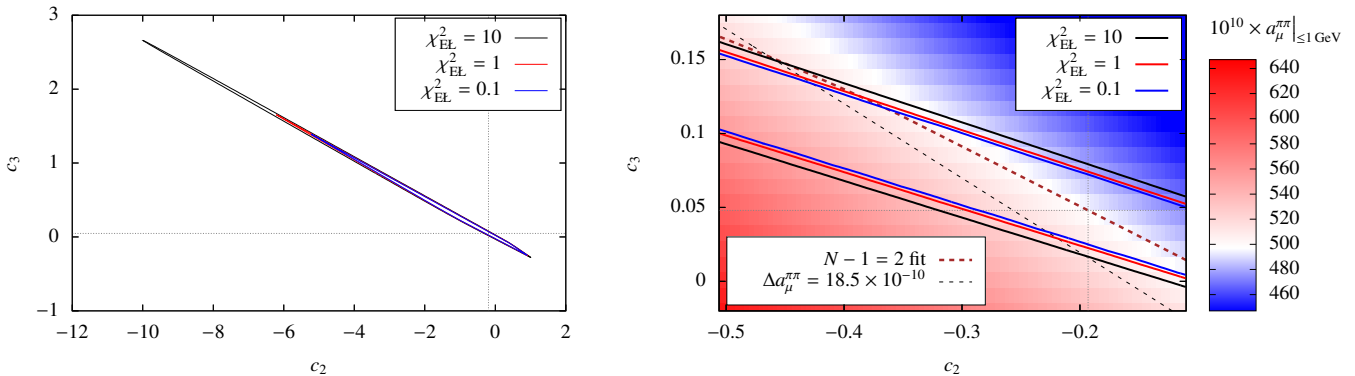

	\centering
	\scalebox{0.75}{
	\input{plots/EL-N-1-2}
	\input{plots/EL-N-1-2-detail}}%
	\caption{Impact of the \EL{} bound on the $\chi^2$ for $N-1=2$ when varying the free parameters $c_2$ and $c_3$ away from the central fit result (denoted by dotted lines). Shown are the regions corresponding to $\chi^2_\text{\EL{}} \in\{ 0.1, 1, 10\}$. The right plot shows in more detail the parameter region of interest and the results for $\amupipi$ as a heat-map overlay. The brown dashed line shows the path of the fit in scenario (2).}
	\label{img:EL2}
\end{figure*}

Starting from the central fit results, we now modify the contribution to $a_\mu^\text{HVP}$ by including in the fit a hypothetical ``lattice'' observation of $\amupipi$ in the form of an additional contribution to the $\chi^2$ function that we minimize. The fit output for $\amupipi$ is then pulled away from the central fit result in Eq.~\eqref{eq:CentralAmu}, depending on the input $\amupipi$ and its uncertainty that acts as a weight. We find it convenient to adopt a tiny uncertainty, because it forces the output for $\amupipi$ to essentially coincide with the input. With a larger uncertainty (i.e., a smaller weight) the fit output for $\amupipi$ will be somewhere between the input and Eq.~\eqref{eq:CentralAmu}. However, the choice of the weights is immaterial because the following studies are all based on the output $\amupipi$. For a given output $\amupipi$, the fit always finds the parameter values that minimize the tension with the cross-section data. We consider the following three scenarios:
\begin{itemize}
	\item[(1)] ``Low-energy'' scenario: we fix all parameters of the dispersive representation of the VFF to the central fit results with $N-1=4$ without ``lattice'' input for $\amupipi$, apart from the two phase-shift parameters $\delta_1^1(s_0)$ and $\delta_1^1(s_1)$, which are used as free parameters in a fit to data and ``lattice'' input for $\amupipi$.
	\item[(2)] ``High-energy'' scenario: we fix all parameters apart from the parameters $c_k$ in the conformal polynomial.
	\item[(3)] Combined scenario: all parameters are used as free fit parameters.
\end{itemize}
We are interested in the region of the parameter space that allows for a significant upward shift in $a_\mu^{\pi\pi}$. For definiteness, we take
\beq
\label{Deltaamu_becnhmark}
	\Delta a_\mu^{\pi\pi} \big|_{\le1\GeV} \lesssim 18.5 \times 10^{-10}
\eeq
as reference point, which corresponds to the difference between Eqs.~\eqref{ee} and~\eqref{eq:LatticeAverage}.

The dependence of the VFF on the two free phase parameters
$\delta_1^1(s_0)$ and $\delta_1^1(s_1)$ is intertwined with the solution of
the Roy equations for the phase $\delta_1^1(s)$, which in turn determines
the Omn\`es function~\eqref{omnes}. In contrast, the dependence on the
parameters in the conformal polynomial is much more direct, as the
constraint that removes the $S$-wave singularity is a linear relation
between the parameters $c_k$. Therefore, the VFF is linear in the
parameters $c_k$ and the same is true for the contribution to the charge
radius, while $a_\mu^{\pi\pi}$ and $\Dalphahadpipi$ are quadratic in the
conformal parameters $c_k$. However, in the relevant parameter range the
non-linearities prove to be very small. 

In order to further restrict possible variants in scenarios (2) and (3), we first investigate the role of the \EL{} bound in the context of variations of $\amupipi$.

\section{Constraints due to the \EL{} bound}

The \EL{} bound~\eqref{eq:ELBoundAlt} provides an additional restriction on
the permissible parameter space that is independent of the two-pion
cross-section measurements. Using the implementation of
Ref.~\cite{Colangelo:2018mtw} and the data compilation of
Ref.~\cite{Eidelman:2003uh}, this constraint leads to a steep rise of the
$\chi^2$ function unless the inelastic phase stays small. To illustrate
this effect, we consider scenario (2) and fit configurations with
$N-1=1\ldots4$ free parameters in the conformal polynomial. Starting from
the central fit results, we vary the input value for $\amupipi$. The impact
of the \EL{} bound on the $\chi^2$ is shown in Fig.~\ref{img:EL1}, as a
function of the fit output $\amupipi$. We find that the bound severely
restricts the possible changes in $a_\mu^{\pi\pi}$ for $N-1=1$: inducing
larger shifts with only a single free parameter in the conformal polynomial
automatically leads to a significant effect in the inelastic phase that
violates the \EL{} bound, thus excluding such a scenario. With two free
parameters in the conformal polynomial, the \EL{} bound permits larger
changes in $a_\mu^{\pi\pi}$, but still imposes a restriction. To evade the
\EL{} bound for large changes in $a_\mu^{\pi\pi}$, more freedom in the
parameterization is required, and indeed the situation changes if we
consider three or more free parameters in the conformal polynomial, see
Fig.~\ref{img:EL1}. 

In order to better understand this effect, we consider in some detail the case of $N-1=2$. The fit to data alone leads to
\beq
	a_\mu^{\pi\pi} \big|_{\le1\GeV}^{N-1=2} = 497.0(1.4) \times 10^{-10}.
\eeq
Varying the two parameters $c_{2,3}$ away from the central fit results, we find that the \EL{} bound gives a contribution to the $\chi^2$ that results in a strong anti-correlation between permissible values for the two free parameters. This is illustrated in Fig.~\ref{img:EL2}, where we show the contours for $\chi^2_\text{\EL{}} \in\{ 0.1, 1, 10\}$ in the $c_2$--$c_3$ plane. In the close-up plot, we also overlay a heat map for the resulting value of $\amupipi$. Accordingly, for two free parameters in the conformal polynomial the \EL{} bound alone no longer excludes very large shifts in $a_\mu^{\pi\pi}$, as shown by the ellipses in Fig.~\ref{img:EL2}. However, large parts of the $\chi^2_\text{\EL{}}$ ellipsis are in strong tension with the cross-section data. Minimizing the total $\chi^2$ in scenario (2) results in the brown dashed path in Fig.~\ref{img:EL2}, which corresponds to the brown curve shown in Fig.~\ref{img:EL1}. For even more free parameters $N-1>2$, the situation remains qualitatively similar:
the \EL{} bound again strongly correlates the free parameters of the conformal polynomial, essentially imposing one linear constraint, but the values of $a_\mu^{\pi\pi}$ that can be reached are no longer bounded. Therefore, in the following we will only consider fit variants with $N-1=3$ and $N-1=4$, where the \EL{} bound is easily fulfilled even for large shifts in $a_\mu^{\pi\pi}$.

\section{Correlations with $\boldsymbol{\Delta\alpha_\text{had}^{(5)}}$ and $\boldsymbol{\langle r_\pi^2\rangle}$}
\label{sec:Correlations}

\begin{figure}[t]
	\centering
	\scalebox{0.8}{
	\input{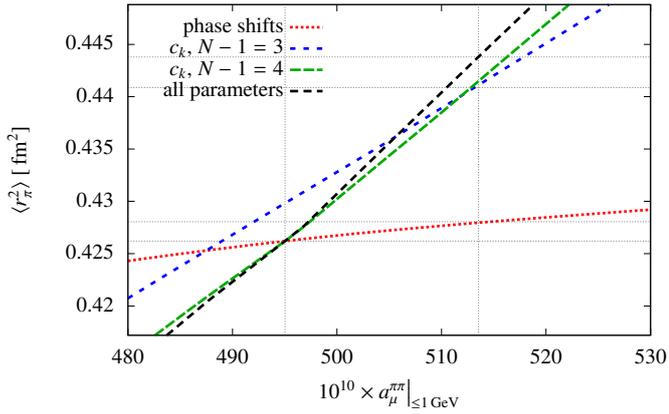}}%
	\caption{Correlations between $a_\mu^{\pi\pi}$ and $\<r_\pi^2\>$ as induced in three different scenarios: a ``low-energy'' scenario (1), where shifts in $a_\mu^{\pi\pi}$ are induced by changes in the phase-shift parameters $\delta_1^1(s_0)$, $\delta_1^1(s_1)$; two ``high-energy'' scenarios (2), where the shifts are due to changes in the conformal polynomial with $N-1=3$ or $N-1=4$; and a combined scenario (3) with $N-1=4$, where all free parameters in the dispersive representation of the pion VFF are allowed to vary.}
	\label{img:CorrelationsRadius}
\end{figure}

\begin{figure}[t]
	\centering
	\scalebox{0.8}{
	\input{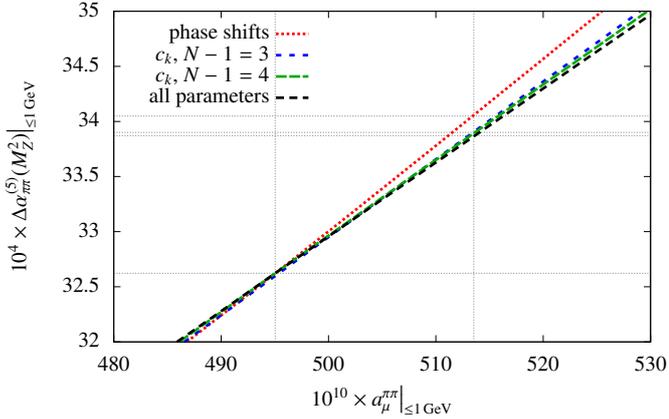}}%
	\caption{Correlations between $a_\mu^{\pi\pi}$ and $\Delta \alpha^{(5)}_{\pi\pi} (M_Z^2)$ for the same scenarios as in Fig.~\ref{img:CorrelationsRadius}.}
	\label{img:CorrelationsAlpha}
\end{figure}

We now turn our attention to the correlations among the three quantities derived from HVP---the two-pion contribution to the anomalous magnetic moment of the muon $a_\mu^{\pi\pi}$, the pion charge radius $\< r_\pi^2 \>$, and the two-pion contribution to the hadronic running of $\alpha$, $\Dalphahadpipi$. We vary the hypothetical ``lattice'' input for $\amupipi$, perform the fits according to the three scenarios defined in Sect.~\ref{sec:ChangingHVP}, and compute the resulting output values for the three quantities. The results in Figs.~\ref{img:CorrelationsRadius} and~\ref{img:CorrelationsAlpha} show the correlations of $a_\mu^{\pi\pi}$ with $\<r_\pi^2\>$ and $\Dalphahadpipi$, respectively, as induced in each of the scenarios. 

\begin{figure}[t]
\centering
	\scalebox{0.8}{
	\input{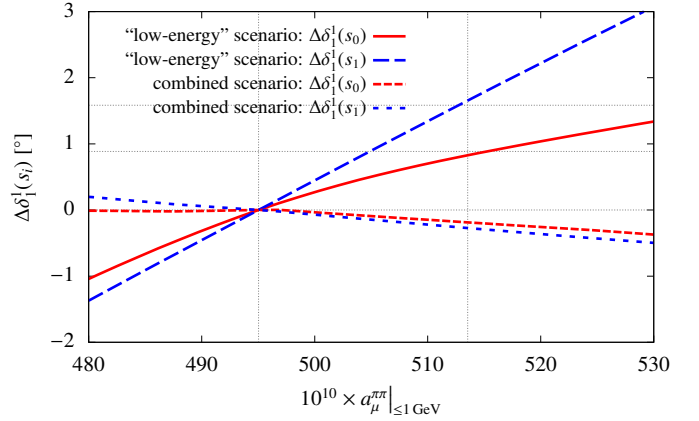}}%
	\caption{Change in the phase shift $\delta_1^1$ at $s_0=(0.8\GeV)^2$ and $s_1=(1.15\GeV)^2$ as a function of $a_\mu^{\pi\pi}$. In scenario (1) only these two parameters are used to achieve the change in $a_\mu^{\pi\pi}$, while in the combined scenario (3) all parameters are changed simultaneously.}
	\label{img:phases}
\end{figure}

If the changes in $\amupipi$ are induced only by variations of the two phase-shift parameters $\delta_1^1(s_0)$ and $\delta_1^1(s_1)$, they have only little impact on the charge radius $\<r_\pi^2\>$, see Fig.~\ref{img:CorrelationsRadius}. 
Hence, in practice changes of $a_\mu^{\pi\pi}$ induced by these parameters cannot be
detected by a precision measurement of $\<r_\pi^2\>$. However, a scenario where the changes in $\amupipi$ are induced by shifts in
the parameters $c_k$ of the conformal polynomial generates large shifts in
$\<r_\pi^2\>$ and could be constrained by additional information on the
charge radius of the pion, at least in 
principle. At present, lattice determinations of the charge
radius~\cite{Feng:2019geu,Wang:2020nbf} have not yet reached the precision
that could exclude these shifts: the current
lattice uncertainties cover the entire plot range in
Fig.~\ref{img:CorrelationsRadius}, but future progress on the determination
of the charge radius could further constrain the allowed parameter range.
Interestingly, the combined scenario (3) where all parameters are allowed to vary leads to the largest effect in the pion charge radius, even slightly larger than the effect in the scenarios (2). By definition, this is the scenario with minimal tension with the cross-section data, but Fig.~\ref{img:CorrelationsRadius} shows that this comes at the expense of the largest shift in the charge radius. 

In contrast to the pion charge radius, all scenarios lead to very similar correlations with the hadronic running of $\alpha$, as shown in Fig.~\ref{img:CorrelationsAlpha}. A shift in $\amupipi$ by $18.5\times10^{-10}$ corresponds to a shift in $\Dalphahadpipi\big|_{\le1\GeV}$ between $1.2 \times 10^{-4}$ and $1.4 \times 10^{-4}$, as shown in Fig.~\ref{img:CorrelationsAlpha}.\footnote{This
shift is slightly smaller than the $1.8\times 10^{-4}$ estimated in
Ref.~\cite{Crivellin:2020zul} if the relative changes occur below
$1.94\GeV$ but are otherwise energy independent. Shifts of this size
violate the bound on $\Delta \alpha^{(5)}_\text{had}(M_Z^2)$ derived in
Ref.~\cite{deRafael:2020uif}. Since this bound was derived on the basis of
assumptions (dim-$6$ operator as sole origin of the shift in $\Delta
\alpha^{(5)}_\text{had}(M_Z^2)$ and an arbitrary scale choice when converting
the derivative of the HVP function to $\Delta
\alpha^{(5)}_\text{had}(M_Z^2)$), we have to conclude that these 
assumptions are not tenable. The result for $\Delta\alpha^{(5)}_\text{had}(M_Z^2)$ indicated by 
Ref.~\cite{Borsanyi:2020mff} leads to the same conclusion.} 
The existence of such a correlation emerges because we do not allow for
arbitrary changes in the hadronic cross section: while in general the two
quantities need not be correlated due to the different energy dependence of
their kernel functions, we find that a correlation does arise if only
changes in the $\pi\pi$ channel are considered as allowed by analyticity
and unitarity constraints, while trying to minimize the tension with the
$\pi\pi$ cross-section data. 

\section{Impact on the phase shift and cross section}

\begin{figure}[t]
	\centering
	\scalebox{0.62}{
	\input{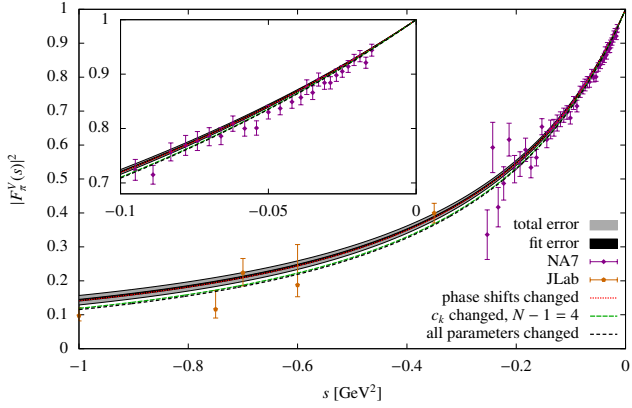}}%
	\caption{Close-up view of the spacelike region. The JLab data~\cite{Horn:2006tm,Tadevosyan:2007yd,Huber:2008id,Blok:2008jy} are not used in the fit.} 
	\label{img:VFFSpacelike}   
\end{figure}

In scenario (1) we only allow the two phase-shift parameters
$\delta_1^1(s_0)$ and $\delta_1^1(s_1)$ to deviate from the central fit
results to data. If only the phase at $s_0=(0.8\GeV)^2$ were varied, a huge
change in the phase shift of about $\Delta\delta_1^1(s_0) = 10^\circ$ would
be necessary to obtain a shift in $\amupipi$ by $18.5\times10^{-10}$. On
the other hand, such a change in $a_\mu^{\pi\pi}$ could be induced by the
parameter $\delta_1^1(s_1)$ alone with a shift by $1.8^\circ$. If we fit
the two parameters simultaneously to a combination of the space- and
time-like data on the VFF and the hypothetical ``lattice'' input on
$\amupipi$, a shift in $\amupipi$ by $18.5\times10^{-10}$ then corresponds
to modest changes in the phase by $\Delta\delta_1^1(s_0) = 0.8^\circ$ and
$\Delta\delta_1^1(s_1) = 1.7^\circ$, see Fig.~\ref{img:phases}. 
We note that the partial-wave solutions given in Eq.~\eqref{PWA} would actually favor values slightly below our reference point~\eqref{pipi_phase_final}, but certainly exclude the required change in $\delta_1^1(s_0)$ if the shift in $\amupipi$ were induced by this parameter alone. 

As discussed in Sect.~\ref{sec:Correlations}, indirect constraints on scenario (1) from a determination of the pion charge radius seem out of reach. However, direct constraints on $\delta_1^1(s_0)$ and $\delta_1^1(s_1)$ could
be obtained from lattice determinations of the elastic $\pi\pi$ phase
shift~\cite{Wilson:2015dqa,Bali:2015gji,Guo:2016zos,Fu:2016itp,Alexandrou:2017mpi,Andersen:2018mau,Werner:2019hxc,Erben:2019nmx,Fischer:2020fvl,Niehus:2020gmf},
not only at these exact points in energy, but in the whole $\rho$ resonance
region: given the phase values $\delta_1^1(s_{0,1})$, the Roy solutions determine the modified phase shift over the whole energy range. However, the precision of lattice data is not yet sufficient
to add meaningful constraints to the parameter space, 
and only a significant increase in precision will have an impact on
$a_\mu^\text{HVP}$ determinations.

Figure~\ref{img:phases} also shows the resulting shifts in the phase parameters for scenario (3), $\Delta\delta_1^1(s_0) = -0.2^\circ$ and $\Delta\delta_1^1(s_1) = -0.3^\circ$. As discussed in Sect.~\ref{sec:Correlations}, it is most promising to indirectly constrain such a scenario with an improved determination of the pion charge radius. In fact, not only the radius is relevant in this regard, but the VFF in the whole space-like region, as shown in Fig.~\ref{img:VFFSpacelike}. Scenarios (2) and (3) move the curve outside the error band of the central fit to data. Precise lattice-QCD determinations of the space-like VFF~\cite{Wang:2020nbf} could start to discriminate between the central solution and these shifted variants. Consistently with the small effect on the radius, scenario (1) with shifts only in the two phase-shift parameters has a negligible effect on the space-like VFF: the shifted solution remains well within the uncertainties of the central fit result.

\begin{figure}[t]
	\centering
	\scalebox{0.7}{
	\input{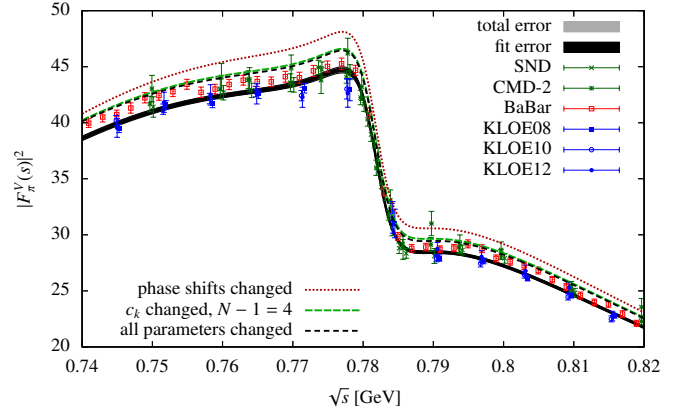}}%
	\caption{Close-up view of the $\rho$--$\omega$ interference region.}
	\label{img:VFFZoom}
\end{figure}

Finally, we take a closer look at the pion VFF in the time-like region. The dispersive representation of the VFF allows us to quantify in detail how the cross sections would need to be altered to achieve a given change in $\amupipi$, in each of the three scenarios. 
In Fig.~\ref{img:VFFZoom}, a close-up view of the $\rho$--$\omega$
interference region is shown. It reveals that if the change in $\amupipi$
were explained with the help of $\delta_1^1(s_{0,1})$, a dramatic shift of
up to $8\%$ of the cross section would be necessary. If the shift were
obtained by changing the parameters $c_k$, the effect in the cross section
at the $\rho$ resonance would be only about half as large, although the
resulting cross section would still lie far outside the combined fit to the
data. The combined scenario is very close to the one where shifts are only
allowed in the parameters $c_k$. 

\begin{figure}[t] 
	\centering
	\scalebox{0.7}{
	\input{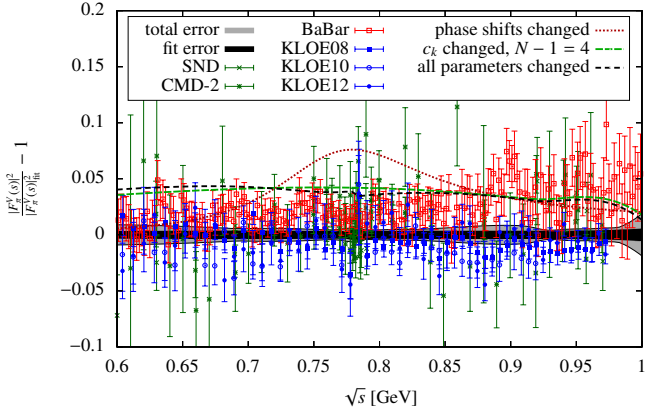}}%
	\caption{Comparison of the data sets and the shifted variants of the VFF, relative to the central fit solution.}
	\label{img:VFFRelative}
\end{figure}

In Fig.~\ref{img:VFFRelative}, we compare both the data sets and the
shifted variants of the VFF to the central fit result, as the relative
differences normalized to the fit result. We again see that by using the
conformal polynomial to induce the shift, the effect on the cross sections
is smaller around the $\rho$ resonance than in the scenario with a shift in
$\delta_1^1(s_{0,1})$, while the effect is larger below about
$0.72\GeV$. Compared to the spread of the data points, the necessary shift
in the cross sections is again significant, although less drastic than in
scenario (1), where the changes are concentrated in the $\rho$ region. This
is consistent with the fact that the conformal polynomial parameterizes the
effects of inelasticities above the $\pi\omega$ threshold. 

While Figs.~\ref{img:VFFZoom} and~\ref{img:VFFRelative} make it 
evident that the changes in the cross section that would generate the
desired change in $\amupipi$ are incompatible with the data,
Fig.~\ref{img:chi2} shows the corresponding change in $\chi^2$ as a
function of $\amupipi$, and provides a quantitative measure of the
discrepancy. The most dramatic clash with the data would be in scenario
(1), but even in the other two any significant change in $\amupipi$ comes
at the price of huge increases in $\chi^2$.
These increases can be compared to the well-known tension between individual $e^+e^-$ data sets. The central fit results of Ref.~\cite{Colangelo:2018mtw} reach a total $\chi^2$ of $776$ with $627$ degrees of freedom. The tension is reflected by an error inflation included in Eq.~\eqref{eq:CentralAmu} of $\sqrt{\chi^2/\mathrm{dof}} = 1.11$. For the target shift of $\Delta\amupipi = 18.5 \times 10^{-10}$, even scenario (3) leads to a total $\chi^2$ of $941$.

The results in Figs.~\ref{img:VFFZoom} and~\ref{img:VFFRelative} show that
to minimize the effect in the cross section, the changes mainly affect the
inelastic part of the VFF parameterization and thus energies above the
$\pi\omega$ threshold. In principle, these inelastic contributions could be
further constrained by $e^+e^-\to2\pi$ data above
$1\GeV$~\cite{Aulchenko:2006na,Aubert:2009ad,Lees:2012cj},
$\tau\to\pi\pi\nu_\tau$~\cite{Fujikawa:2008ma}, and explicit input on the
inelastic channels, but this requires an extension of our dispersive
formalism that will be left for future work. We remark that any changes in
the physics above $1\GeV$ will also have an impact on
$\Dalphahadpipi$, which is not yet accounted for here: the higher
in energy these changes are pushed, the higher the risk to exacerbate
tensions in the global electroweak
fit~\cite{Passera:2008jk,Crivellin:2020zul,Keshavarzi:2020bfy,Malaescu:2020zuc}.

\section{Conclusions}

In this Letter we examined the two-pion contribution to HVP in view of
recent hints from lattice-QCD calculations that its contribution to the
anomalous magnetic moment of the muon could be much larger than obtained
from $e^+e^-\to\text{hadrons}$ cross-section data, with most of the changes
concentrated at low energies. We relied on a dispersive representation of
the pion vector form factor and studied which of its parameters could be
varied without contradicting other low-energy observables besides the
$e^+e^-\to 2\pi$ cross section itself. We identified three scenarios: (1) 
where only the elastic $\pi \pi$ phase shift, or (2) where only inelastic
effects, or (3) all parameters at the same time are allowed to change, see
Sect.~\ref{sec:ChangingHVP} for more details.  
In these scenarios, we then derived the correlations with
the pion charge radius and the hadronic running of the fine-structure
constant.

\begin{figure}[t]
	\centering
	\scalebox{0.8}{
	\input{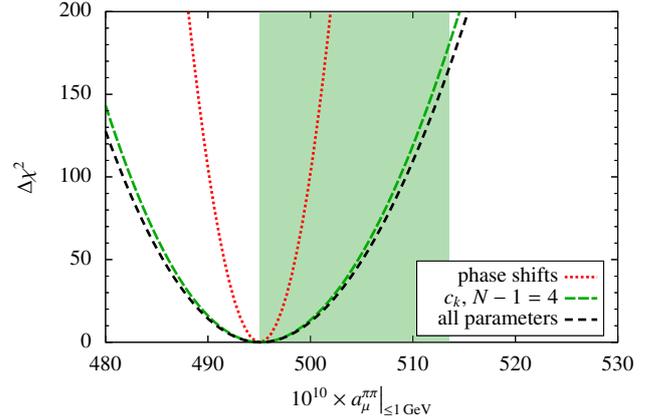}}%
	\caption{Increase in the $\chi^2$ as a function of the fit output $\amupipi$ in the three scenarios, excluding the contribution of the ``lattice'' input (since this depends on the arbitrary uncertainty that acts as a weight, see Sect.~\ref{sec:ChangingHVP}).}
	\label{img:chi2}
\end{figure}

We found that in scenario (1) the changes in the cross section are mainly
concentrated around the $\rho$ resonance, amounting to a relative effect of
up to $8\%$, see Figs.~\ref{img:VFFZoom} and \ref{img:VFFRelative}, while
in scenarios (2) and (3) the changes are more uniformly distributed over
the entire energy range, at a level around $4\%$. The first insight from our analysis is thus that a largely uniform change in the cross section is actually allowed by the constraints from analyticity, unitarity, as well as low-energy hadron phenomenology. Moreover, this is the configuration that minimizes the discrepancy with
the data as one tries to increase $\amupipi$ while respecting all constraints, but still even this scenario 
is in strong disagreement with the $e^+e^-\to 2\pi$
data, see Fig.~\ref{img:chi2}.

The correlations with the pion charge radius and the hadronic running of
the fine-structure constant are shown in Figs.~\ref{img:CorrelationsRadius}
and \ref{img:CorrelationsAlpha}, respectively. One of our main conclusions
is that in our framework we can establish a firm correlation between
$\amupipi$ and $\Dalphahadpipi$: the required change in the former implies
an upward shift between $1.2\times 10^{-4}$ and $1.4\times 10^{-4}$ in the
latter for all scenarios. For the charge radius the correlation with
$\amupipi$ depends on the scenario, with the largest effect arising in
scenario (3), the one for which the change in the cross section is
minimized.  A similar observation applies to the entire space-like region,
see Fig.~\ref{img:VFFSpacelike}. This opens the possibility to challenge
this scenario with future lattice-QCD calculations of the pion charge
radius as well as the space-like pion form
factor~\cite{Feng:2019geu,Wang:2020nbf}. Competitive constraints would
require a precision around $\Delta\langle r^2_\pi\rangle=0.005\fm^2$, a
factor $3$ below the sensitivity of Ref.~\cite{Wang:2020nbf}.  Similarly, a
precision calculation of the $P$-wave $\pi\pi$ phase shift would provide
further independent constraints on our dispersive representation, but here
the precision goal of $\Delta \delta^1_1(s_{0,1})=2^\circ$ would require
significant advances over current calculations.

To further improve the phenomenological determination of the two-pion
contribution to HVP, the most important future development
naturally concerns new $e^+e^-\to 2\pi$ data, with
\mbox{BESIII}~\cite{Ablikim:2015orh,Ablikim:2020bah} and
SND~\cite{Achasov:2020iys} supporting the results already included in the
present analysis, and new data from \mbox{CMD-3}~\cite{Ignatov:2019omb}
forthcoming. As for direct lattice-QCD evaluations of the HVP
contribution, the results of Ref.~\cite{Borsanyi:2020mff} are being
scrutinized by other lattice collaborations, and more detailed comparisons
to phenomenology will allow for refined conclusions as to where the
$e^+e^-\to\text{hadrons}$ cross section would need to be modified. In
addition, a direct measurement of HVP in the space-like region would become
possible with the MUonE project~\cite{Abbiendi:2016xup,Banerjee:2020tdt},
providing further complementary information on the role of HVP in the SM
prediction for the anomalous magnetic moment of the muon.

\section*{Acknowledgments}

Financial support by  the DOE (Grant No.\ DE-SC0009919) and the Swiss
National Science Foundation (Project No.\ PCEFP2\_181117 and Grant No.\
200020\_175791) is gratefully acknowledged.    
  
\bibliographystyle{apsrev4-1_mod}
\balance
\biboptions{sort&compress}
\bibliography{AMM}

\begin{thebibliography}{110}%
\makeatletter
\providecommand \@ifxundefined [1]{%
 \@ifx{#1\undefined}
}%
\providecommand \@ifnum [1]{%
 \ifnum #1\expandafter \@firstoftwo
 \else \expandafter \@secondoftwo
 \fi
}%
\providecommand \@ifx [1]{%
 \ifx #1\expandafter \@firstoftwo
 \else \expandafter \@secondoftwo
 \fi
}%
\providecommand \natexlab [1]{#1}%
\providecommand \enquote  [1]{``#1''}%
\providecommand \bibnamefont  [1]{#1}%
\providecommand \bibfnamefont [1]{#1}%
\providecommand \citenamefont [1]{#1}%
\providecommand \href@noop [0]{\@secondoftwo}%
\providecommand \href [0]{\begingroup \@sanitize@url \@href}%
\providecommand \@href[1]{\@@startlink{#1}\@@href}%
\providecommand \@@href[1]{\endgroup#1\@@endlink}%
\providecommand \@sanitize@url [0]{\catcode `\\12\catcode `\$12\catcode
  `\&12\catcode `\#12\catcode `\^12\catcode `\_12\catcode `\%12\relax}%
\providecommand \@@startlink[1]{}%
\providecommand \@@endlink[0]{}%
\providecommand \url  [0]{\begingroup\@sanitize@url \@url }%
\providecommand \@url [1]{\endgroup\@href {#1}{\urlprefix }}%
\providecommand \urlprefix  [0]{URL }%
\providecommand \Eprint [0]{\href }%
\providecommand \doibase [0]{http://dx.doi.org/}%
\providecommand \selectlanguage [0]{\@gobble}%
\providecommand \bibinfo  [0]{\@secondoftwo}%
\providecommand \bibfield  [0]{\@secondoftwo}%
\providecommand \translation [1]{[#1]}%
\providecommand \BibitemOpen [0]{}%
\providecommand \bibitemStop [0]{}%
\providecommand \bibitemNoStop [0]{.\EOS\space}%
\providecommand \EOS [0]{\spacefactor3000\relax}%
\providecommand \BibitemShut  [1]{\csname bibitem#1\endcsname}%
\let\auto@bib@innerbib\@empty
\bibitem [{\citenamefont {Aoyama}\ \emph {et~al.}(2020)\citenamefont {Aoyama}
  \emph {et~al.}}]{Aoyama:2020ynm}%
  \BibitemOpen
  \bibfield  {author} {\bibinfo {author} {\bibfnamefont {T.}~\bibnamefont
  {Aoyama}} \emph {et~al.},\ }\href {\doibase 10.1016/j.physrep.2020.07.006}
  {\bibfield  {journal} {\bibinfo  {journal} {Phys. Rept.}\ }\textbf {\bibinfo
  {volume} {887}},\ \bibinfo {pages} {1} (\bibinfo {year} {2020})},\ \Eprint
  {http://arxiv.org/abs/2006.04822} {arXiv:2006.04822 [hep-ph]}\BibitemShut
  {NoStop}%
\bibitem [{\citenamefont {Aoyama}\ \emph {et~al.}(2012)\citenamefont {Aoyama},
  \citenamefont {Hayakawa}, \citenamefont {Kinoshita},\ and\ \citenamefont
  {Nio}}]{Aoyama:2012wk}%
  \BibitemOpen
  \bibfield  {author} {\bibinfo {author} {\bibfnamefont {T.}~\bibnamefont
  {Aoyama}}, \bibinfo {author} {\bibfnamefont {M.}~\bibnamefont {Hayakawa}},
  \bibinfo {author} {\bibfnamefont {T.}~\bibnamefont {Kinoshita}}, \ and\
  \bibinfo {author} {\bibfnamefont {M.}~\bibnamefont {Nio}},\ }\href {\doibase
  10.1103/PhysRevLett.109.111808} {\bibfield  {journal} {\bibinfo  {journal}
  {Phys. Rev. Lett.}\ }\textbf {\bibinfo {volume} {109}},\ \bibinfo {pages}
  {111808} (\bibinfo {year} {2012})},\ \Eprint {http://arxiv.org/abs/1205.5370}
  {arXiv:1205.5370 [hep-ph]}\BibitemShut {NoStop}%
\bibitem [{\citenamefont {Aoyama}\ \emph {et~al.}(2019)\citenamefont {Aoyama},
  \citenamefont {Kinoshita},\ and\ \citenamefont {Nio}}]{Aoyama:2019ryr}%
  \BibitemOpen
  \bibfield  {author} {\bibinfo {author} {\bibfnamefont {T.}~\bibnamefont
  {Aoyama}}, \bibinfo {author} {\bibfnamefont {T.}~\bibnamefont {Kinoshita}}, \
  and\ \bibinfo {author} {\bibfnamefont {M.}~\bibnamefont {Nio}},\ }\href
  {\doibase 10.3390/atoms7010028} {\bibfield  {journal} {\bibinfo  {journal}
  {Atoms}\ }\textbf {\bibinfo {volume} {7}},\ \bibinfo {pages} {28} (\bibinfo
  {year} {2019})}\BibitemShut {NoStop}%
\bibitem [{\citenamefont {Czarnecki}\ \emph {et~al.}(2003)\citenamefont
  {Czarnecki}, \citenamefont {Marciano},\ and\ \citenamefont
  {Vainshtein}}]{Czarnecki:2002nt}%
  \BibitemOpen
  \bibfield  {author} {\bibinfo {author} {\bibfnamefont {A.}~\bibnamefont
  {Czarnecki}}, \bibinfo {author} {\bibfnamefont {W.~J.}\ \bibnamefont
  {Marciano}}, \ and\ \bibinfo {author} {\bibfnamefont {A.}~\bibnamefont
  {Vainshtein}},\ }\href {\doibase 10.1103/PhysRevD.67.073006} {\bibfield
  {journal} {\bibinfo  {journal} {Phys. Rev.}\ }\textbf {\bibinfo {volume}
  {D67}},\ \bibinfo {pages} {073006} (\bibinfo {year} {2003})},\ \bibinfo
  {note} {[Erratum: Phys. Rev. {\bf D73}, 119901 (2006)]},\ \Eprint
  {http://arxiv.org/abs/hep-ph/0212229} {arXiv:hep-ph/0212229
  [hep-ph]}\BibitemShut {NoStop}%
\bibitem [{\citenamefont {Gnendiger}\ \emph {et~al.}(2013)\citenamefont
  {Gnendiger}, \citenamefont {St{\"o}ckinger},\ and\ \citenamefont
  {St{\"o}ckinger-Kim}}]{Gnendiger:2013pva}%
  \BibitemOpen
  \bibfield  {author} {\bibinfo {author} {\bibfnamefont {C.}~\bibnamefont
  {Gnendiger}}, \bibinfo {author} {\bibfnamefont {D.}~\bibnamefont
  {St{\"o}ckinger}}, \ and\ \bibinfo {author} {\bibfnamefont {H.}~\bibnamefont
  {St{\"o}ckinger-Kim}},\ }\href {\doibase 10.1103/PhysRevD.88.053005}
  {\bibfield  {journal} {\bibinfo  {journal} {Phys. Rev.}\ }\textbf {\bibinfo
  {volume} {D88}},\ \bibinfo {pages} {053005} (\bibinfo {year} {2013})},\
  \Eprint {http://arxiv.org/abs/1306.5546} {arXiv:1306.5546
  [hep-ph]}\BibitemShut {NoStop}%
\bibitem [{\citenamefont {Davier}\ \emph {et~al.}(2017)\citenamefont {Davier},
  \citenamefont {Hoecker}, \citenamefont {Malaescu},\ and\ \citenamefont
  {Zhang}}]{Davier:2017zfy}%
  \BibitemOpen
  \bibfield  {author} {\bibinfo {author} {\bibfnamefont {M.}~\bibnamefont
  {Davier}}, \bibinfo {author} {\bibfnamefont {A.}~\bibnamefont {Hoecker}},
  \bibinfo {author} {\bibfnamefont {B.}~\bibnamefont {Malaescu}}, \ and\
  \bibinfo {author} {\bibfnamefont {Z.}~\bibnamefont {Zhang}},\ }\href
  {\doibase 10.1140/epjc/s10052-017-5161-6} {\bibfield  {journal} {\bibinfo
  {journal} {Eur. Phys. J.}\ }\textbf {\bibinfo {volume} {C77}},\ \bibinfo
  {pages} {827} (\bibinfo {year} {2017})},\ \Eprint
  {http://arxiv.org/abs/1706.09436} {arXiv:1706.09436 [hep-ph]}\BibitemShut
  {NoStop}%
\bibitem [{\citenamefont {Keshavarzi}\ \emph {et~al.}(2018)\citenamefont
  {Keshavarzi}, \citenamefont {Nomura},\ and\ \citenamefont
  {Teubner}}]{Keshavarzi:2018mgv}%
  \BibitemOpen
  \bibfield  {author} {\bibinfo {author} {\bibfnamefont {A.}~\bibnamefont
  {Keshavarzi}}, \bibinfo {author} {\bibfnamefont {D.}~\bibnamefont {Nomura}},
  \ and\ \bibinfo {author} {\bibfnamefont {T.}~\bibnamefont {Teubner}},\ }\href
  {\doibase 10.1103/PhysRevD.97.114025} {\bibfield  {journal} {\bibinfo
  {journal} {Phys. Rev.}\ }\textbf {\bibinfo {volume} {D97}},\ \bibinfo {pages}
  {114025} (\bibinfo {year} {2018})},\ \Eprint
  {http://arxiv.org/abs/1802.02995} {arXiv:1802.02995 [hep-ph]}\BibitemShut
  {NoStop}%
\bibitem [{\citenamefont {Colangelo}\ \emph {et~al.}(2019)\citenamefont
  {Colangelo}, \citenamefont {Hoferichter},\ and\ \citenamefont
  {Stoffer}}]{Colangelo:2018mtw}%
  \BibitemOpen
  \bibfield  {author} {\bibinfo {author} {\bibfnamefont {G.}~\bibnamefont
  {Colangelo}}, \bibinfo {author} {\bibfnamefont {M.}~\bibnamefont
  {Hoferichter}}, \ and\ \bibinfo {author} {\bibfnamefont {P.}~\bibnamefont
  {Stoffer}},\ }\href {\doibase 10.1007/JHEP02(2019)006} {\bibfield  {journal}
  {\bibinfo  {journal} {JHEP}\ }\textbf {\bibinfo {volume} {02}},\ \bibinfo
  {pages} {006} (\bibinfo {year} {2019})},\ \Eprint
  {http://arxiv.org/abs/1810.00007} {arXiv:1810.00007 [hep-ph]}\BibitemShut
  {NoStop}%
\bibitem [{\citenamefont {Hoferichter}\ \emph {et~al.}(2019)\citenamefont
  {Hoferichter}, \citenamefont {Hoid},\ and\ \citenamefont
  {Kubis}}]{Hoferichter:2019gzf}%
  \BibitemOpen
  \bibfield  {author} {\bibinfo {author} {\bibfnamefont {M.}~\bibnamefont
  {Hoferichter}}, \bibinfo {author} {\bibfnamefont {B.-L.}\ \bibnamefont
  {Hoid}}, \ and\ \bibinfo {author} {\bibfnamefont {B.}~\bibnamefont {Kubis}},\
  }\href {\doibase 10.1007/JHEP08(2019)137} {\bibfield  {journal} {\bibinfo
  {journal} {JHEP}\ }\textbf {\bibinfo {volume} {08}},\ \bibinfo {pages} {137}
  (\bibinfo {year} {2019})},\ \Eprint {http://arxiv.org/abs/1907.01556}
  {arXiv:1907.01556 [hep-ph]}\BibitemShut {NoStop}%
\bibitem [{\citenamefont {Davier}\ \emph {et~al.}(2020)\citenamefont {Davier},
  \citenamefont {Hoecker}, \citenamefont {Malaescu},\ and\ \citenamefont
  {Zhang}}]{Davier:2019can}%
  \BibitemOpen
  \bibfield  {author} {\bibinfo {author} {\bibfnamefont {M.}~\bibnamefont
  {Davier}}, \bibinfo {author} {\bibfnamefont {A.}~\bibnamefont {Hoecker}},
  \bibinfo {author} {\bibfnamefont {B.}~\bibnamefont {Malaescu}}, \ and\
  \bibinfo {author} {\bibfnamefont {Z.}~\bibnamefont {Zhang}},\ }\href
  {\doibase 10.1140/epjc/s10052-020-7792-2} {\bibfield  {journal} {\bibinfo
  {journal} {Eur. Phys. J.}\ }\textbf {\bibinfo {volume} {C80}},\ \bibinfo
  {pages} {241} (\bibinfo {year} {2020})},\ \bibinfo {note} {[Erratum: Eur.
  Phys. J. {\bf C80}, 410 (2020)]},\ \Eprint {http://arxiv.org/abs/1908.00921}
  {arXiv:1908.00921 [hep-ph]}\BibitemShut {NoStop}%
\bibitem [{\citenamefont {Keshavarzi}\ \emph
  {et~al.}(2020{\natexlab{a}})\citenamefont {Keshavarzi}, \citenamefont
  {Nomura},\ and\ \citenamefont {Teubner}}]{Keshavarzi:2019abf}%
  \BibitemOpen
  \bibfield  {author} {\bibinfo {author} {\bibfnamefont {A.}~\bibnamefont
  {Keshavarzi}}, \bibinfo {author} {\bibfnamefont {D.}~\bibnamefont {Nomura}},
  \ and\ \bibinfo {author} {\bibfnamefont {T.}~\bibnamefont {Teubner}},\ }\href
  {\doibase 10.1103/PhysRevD.101.014029} {\bibfield  {journal} {\bibinfo
  {journal} {Phys. Rev.}\ }\textbf {\bibinfo {volume} {D101}},\ \bibinfo
  {pages} {014029} (\bibinfo {year} {2020}{\natexlab{a}})},\ \Eprint
  {http://arxiv.org/abs/1911.00367} {arXiv:1911.00367 [hep-ph]}\BibitemShut
  {NoStop}%
\bibitem [{\citenamefont {Hoid}\ \emph {et~al.}(2020)\citenamefont {Hoid},
  \citenamefont {Hoferichter},\ and\ \citenamefont {Kubis}}]{Hoid:2020xjs}%
  \BibitemOpen
  \bibfield  {author} {\bibinfo {author} {\bibfnamefont {B.-L.}\ \bibnamefont
  {Hoid}}, \bibinfo {author} {\bibfnamefont {M.}~\bibnamefont {Hoferichter}}, \
  and\ \bibinfo {author} {\bibfnamefont {B.}~\bibnamefont {Kubis}},\
  }\href@noop {} {\bibfield  {journal} {\bibinfo  {journal} {Eur. Phys. J. C}\
  }\textbf {\bibinfo {volume} {80}},\ \bibinfo {pages} {988} (\bibinfo {year}
  {2020})},\ \Eprint {http://arxiv.org/abs/2007.12696} {arXiv:2007.12696
  [hep-ph]}\BibitemShut {NoStop}%
\bibitem [{\citenamefont {Kurz}\ \emph {et~al.}(2014)\citenamefont {Kurz},
  \citenamefont {Liu}, \citenamefont {Marquard},\ and\ \citenamefont
  {Steinhauser}}]{Kurz:2014wya}%
  \BibitemOpen
  \bibfield  {author} {\bibinfo {author} {\bibfnamefont {A.}~\bibnamefont
  {Kurz}}, \bibinfo {author} {\bibfnamefont {T.}~\bibnamefont {Liu}}, \bibinfo
  {author} {\bibfnamefont {P.}~\bibnamefont {Marquard}}, \ and\ \bibinfo
  {author} {\bibfnamefont {M.}~\bibnamefont {Steinhauser}},\ }\href {\doibase
  10.1016/j.physletb.2014.05.043} {\bibfield  {journal} {\bibinfo  {journal}
  {Phys. Lett.}\ }\textbf {\bibinfo {volume} {B734}},\ \bibinfo {pages} {144}
  (\bibinfo {year} {2014})},\ \Eprint {http://arxiv.org/abs/1403.6400}
  {arXiv:1403.6400 [hep-ph]}\BibitemShut {NoStop}%
\bibitem [{\citenamefont {Melnikov}\ and\ \citenamefont
  {Vainshtein}(2004)}]{Melnikov:2003xd}%
  \BibitemOpen
  \bibfield  {author} {\bibinfo {author} {\bibfnamefont {K.}~\bibnamefont
  {Melnikov}}\ and\ \bibinfo {author} {\bibfnamefont {A.}~\bibnamefont
  {Vainshtein}},\ }\href {\doibase 10.1103/PhysRevD.70.113006} {\bibfield
  {journal} {\bibinfo  {journal} {Phys. Rev.}\ }\textbf {\bibinfo {volume}
  {D70}},\ \bibinfo {pages} {113006} (\bibinfo {year} {2004})},\ \Eprint
  {http://arxiv.org/abs/hep-ph/0312226} {arXiv:hep-ph/0312226
  [hep-ph]}\BibitemShut {NoStop}%
\bibitem [{\citenamefont {Colangelo}\ \emph
  {et~al.}(2014{\natexlab{a}})\citenamefont {Colangelo}, \citenamefont
  {Hoferichter}, \citenamefont {Procura},\ and\ \citenamefont
  {Stoffer}}]{Colangelo:2014dfa}%
  \BibitemOpen
  \bibfield  {author} {\bibinfo {author} {\bibfnamefont {G.}~\bibnamefont
  {Colangelo}}, \bibinfo {author} {\bibfnamefont {M.}~\bibnamefont
  {Hoferichter}}, \bibinfo {author} {\bibfnamefont {M.}~\bibnamefont
  {Procura}}, \ and\ \bibinfo {author} {\bibfnamefont {P.}~\bibnamefont
  {Stoffer}},\ }\href {\doibase 10.1007/JHEP09(2014)091} {\bibfield  {journal}
  {\bibinfo  {journal} {JHEP}\ }\textbf {\bibinfo {volume} {09}},\ \bibinfo
  {pages} {091} (\bibinfo {year} {2014}{\natexlab{a}})},\ \Eprint
  {http://arxiv.org/abs/1402.7081} {arXiv:1402.7081 [hep-ph]}\BibitemShut
  {NoStop}%
\bibitem [{\citenamefont {Colangelo}\ \emph
  {et~al.}(2014{\natexlab{b}})\citenamefont {Colangelo}, \citenamefont
  {Hoferichter}, \citenamefont {Kubis}, \citenamefont {Procura},\ and\
  \citenamefont {Stoffer}}]{Colangelo:2014pva}%
  \BibitemOpen
  \bibfield  {author} {\bibinfo {author} {\bibfnamefont {G.}~\bibnamefont
  {Colangelo}}, \bibinfo {author} {\bibfnamefont {M.}~\bibnamefont
  {Hoferichter}}, \bibinfo {author} {\bibfnamefont {B.}~\bibnamefont {Kubis}},
  \bibinfo {author} {\bibfnamefont {M.}~\bibnamefont {Procura}}, \ and\
  \bibinfo {author} {\bibfnamefont {P.}~\bibnamefont {Stoffer}},\ }\href
  {\doibase 10.1016/j.physletb.2014.09.021} {\bibfield  {journal} {\bibinfo
  {journal} {Phys. Lett.}\ }\textbf {\bibinfo {volume} {B738}},\ \bibinfo
  {pages} {6} (\bibinfo {year} {2014}{\natexlab{b}})},\ \Eprint
  {http://arxiv.org/abs/1408.2517} {arXiv:1408.2517 [hep-ph]}\BibitemShut
  {NoStop}%
\bibitem [{\citenamefont {Colangelo}\ \emph
  {et~al.}(2015{\natexlab{a}})\citenamefont {Colangelo}, \citenamefont
  {Hoferichter}, \citenamefont {Procura},\ and\ \citenamefont
  {Stoffer}}]{Colangelo:2015ama}%
  \BibitemOpen
  \bibfield  {author} {\bibinfo {author} {\bibfnamefont {G.}~\bibnamefont
  {Colangelo}}, \bibinfo {author} {\bibfnamefont {M.}~\bibnamefont
  {Hoferichter}}, \bibinfo {author} {\bibfnamefont {M.}~\bibnamefont
  {Procura}}, \ and\ \bibinfo {author} {\bibfnamefont {P.}~\bibnamefont
  {Stoffer}},\ }\href {\doibase 10.1007/JHEP09(2015)074} {\bibfield  {journal}
  {\bibinfo  {journal} {JHEP}\ }\textbf {\bibinfo {volume} {09}},\ \bibinfo
  {pages} {074} (\bibinfo {year} {2015}{\natexlab{a}})},\ \Eprint
  {http://arxiv.org/abs/1506.01386} {arXiv:1506.01386 [hep-ph]}\BibitemShut
  {NoStop}%
\bibitem [{\citenamefont {Masjuan}\ and\ \citenamefont
  {S{\'a}nchez-Puertas}(2017)}]{Masjuan:2017tvw}%
  \BibitemOpen
  \bibfield  {author} {\bibinfo {author} {\bibfnamefont {P.}~\bibnamefont
  {Masjuan}}\ and\ \bibinfo {author} {\bibfnamefont {P.}~\bibnamefont
  {S{\'a}nchez-Puertas}},\ }\href {\doibase 10.1103/PhysRevD.95.054026}
  {\bibfield  {journal} {\bibinfo  {journal} {Phys. Rev.}\ }\textbf {\bibinfo
  {volume} {D95}},\ \bibinfo {pages} {054026} (\bibinfo {year} {2017})},\
  \Eprint {http://arxiv.org/abs/1701.05829} {arXiv:1701.05829
  [hep-ph]}\BibitemShut {NoStop}%
\bibitem [{\citenamefont {Colangelo}\ \emph
  {et~al.}(2017{\natexlab{a}})\citenamefont {Colangelo}, \citenamefont
  {Hoferichter}, \citenamefont {Procura},\ and\ \citenamefont
  {Stoffer}}]{Colangelo:2017qdm}%
  \BibitemOpen
  \bibfield  {author} {\bibinfo {author} {\bibfnamefont {G.}~\bibnamefont
  {Colangelo}}, \bibinfo {author} {\bibfnamefont {M.}~\bibnamefont
  {Hoferichter}}, \bibinfo {author} {\bibfnamefont {M.}~\bibnamefont
  {Procura}}, \ and\ \bibinfo {author} {\bibfnamefont {P.}~\bibnamefont
  {Stoffer}},\ }\href {\doibase 10.1103/PhysRevLett.118.232001} {\bibfield
  {journal} {\bibinfo  {journal} {Phys. Rev. Lett.}\ }\textbf {\bibinfo
  {volume} {118}},\ \bibinfo {pages} {232001} (\bibinfo {year}
  {2017}{\natexlab{a}})},\ \Eprint {http://arxiv.org/abs/1701.06554}
  {arXiv:1701.06554 [hep-ph]}\BibitemShut {NoStop}%
\bibitem [{\citenamefont {Colangelo}\ \emph
  {et~al.}(2017{\natexlab{b}})\citenamefont {Colangelo}, \citenamefont
  {Hoferichter}, \citenamefont {Procura},\ and\ \citenamefont
  {Stoffer}}]{Colangelo:2017fiz}%
  \BibitemOpen
  \bibfield  {author} {\bibinfo {author} {\bibfnamefont {G.}~\bibnamefont
  {Colangelo}}, \bibinfo {author} {\bibfnamefont {M.}~\bibnamefont
  {Hoferichter}}, \bibinfo {author} {\bibfnamefont {M.}~\bibnamefont
  {Procura}}, \ and\ \bibinfo {author} {\bibfnamefont {P.}~\bibnamefont
  {Stoffer}},\ }\href {\doibase 10.1007/JHEP04(2017)161} {\bibfield  {journal}
  {\bibinfo  {journal} {JHEP}\ }\textbf {\bibinfo {volume} {04}},\ \bibinfo
  {pages} {161} (\bibinfo {year} {2017}{\natexlab{b}})},\ \Eprint
  {http://arxiv.org/abs/1702.07347} {arXiv:1702.07347 [hep-ph]}\BibitemShut
  {NoStop}%
\bibitem [{\citenamefont {Hoferichter}\ \emph
  {et~al.}(2018{\natexlab{a}})\citenamefont {Hoferichter}, \citenamefont
  {Hoid}, \citenamefont {Kubis}, \citenamefont {Leupold},\ and\ \citenamefont
  {Schneider}}]{Hoferichter:2018dmo}%
  \BibitemOpen
  \bibfield  {author} {\bibinfo {author} {\bibfnamefont {M.}~\bibnamefont
  {Hoferichter}}, \bibinfo {author} {\bibfnamefont {B.-L.}\ \bibnamefont
  {Hoid}}, \bibinfo {author} {\bibfnamefont {B.}~\bibnamefont {Kubis}},
  \bibinfo {author} {\bibfnamefont {S.}~\bibnamefont {Leupold}}, \ and\
  \bibinfo {author} {\bibfnamefont {S.~P.}\ \bibnamefont {Schneider}},\ }\href
  {\doibase 10.1103/PhysRevLett.121.112002} {\bibfield  {journal} {\bibinfo
  {journal} {Phys. Rev. Lett.}\ }\textbf {\bibinfo {volume} {121}},\ \bibinfo
  {pages} {112002} (\bibinfo {year} {2018}{\natexlab{a}})},\ \Eprint
  {http://arxiv.org/abs/1805.01471} {arXiv:1805.01471 [hep-ph]}\BibitemShut
  {NoStop}%
\bibitem [{\citenamefont {Hoferichter}\ \emph
  {et~al.}(2018{\natexlab{b}})\citenamefont {Hoferichter}, \citenamefont
  {Hoid}, \citenamefont {Kubis}, \citenamefont {Leupold},\ and\ \citenamefont
  {Schneider}}]{Hoferichter:2018kwz}%
  \BibitemOpen
  \bibfield  {author} {\bibinfo {author} {\bibfnamefont {M.}~\bibnamefont
  {Hoferichter}}, \bibinfo {author} {\bibfnamefont {B.-L.}\ \bibnamefont
  {Hoid}}, \bibinfo {author} {\bibfnamefont {B.}~\bibnamefont {Kubis}},
  \bibinfo {author} {\bibfnamefont {S.}~\bibnamefont {Leupold}}, \ and\
  \bibinfo {author} {\bibfnamefont {S.~P.}\ \bibnamefont {Schneider}},\ }\href
  {\doibase 10.1007/JHEP10(2018)141} {\bibfield  {journal} {\bibinfo  {journal}
  {JHEP}\ }\textbf {\bibinfo {volume} {10}},\ \bibinfo {pages} {141} (\bibinfo
  {year} {2018}{\natexlab{b}})},\ \Eprint {http://arxiv.org/abs/1808.04823}
  {arXiv:1808.04823 [hep-ph]}\BibitemShut {NoStop}%
\bibitem [{\citenamefont {G{\'e}rardin}\ \emph {et~al.}(2019)\citenamefont
  {G{\'e}rardin}, \citenamefont {Meyer},\ and\ \citenamefont
  {Nyffeler}}]{Gerardin:2019vio}%
  \BibitemOpen
  \bibfield  {author} {\bibinfo {author} {\bibfnamefont {A.}~\bibnamefont
  {G{\'e}rardin}}, \bibinfo {author} {\bibfnamefont {H.~B.}\ \bibnamefont
  {Meyer}}, \ and\ \bibinfo {author} {\bibfnamefont {A.}~\bibnamefont
  {Nyffeler}},\ }\href {\doibase 10.1103/PhysRevD.100.034520} {\bibfield
  {journal} {\bibinfo  {journal} {Phys. Rev.}\ }\textbf {\bibinfo {volume}
  {D100}},\ \bibinfo {pages} {034520} (\bibinfo {year} {2019})},\ \Eprint
  {http://arxiv.org/abs/1903.09471} {arXiv:1903.09471 [hep-lat]}\BibitemShut
  {NoStop}%
\bibitem [{\citenamefont {Bijnens}\ \emph {et~al.}(2019)\citenamefont
  {Bijnens}, \citenamefont {Hermansson-Truedsson},\ and\ \citenamefont
  {Rodr{\'i}guez-S{\'a}nchez}}]{Bijnens:2019ghy}%
  \BibitemOpen
  \bibfield  {author} {\bibinfo {author} {\bibfnamefont {J.}~\bibnamefont
  {Bijnens}}, \bibinfo {author} {\bibfnamefont {N.}~\bibnamefont
  {Hermansson-Truedsson}}, \ and\ \bibinfo {author} {\bibfnamefont
  {A.}~\bibnamefont {Rodr{\'i}guez-S{\'a}nchez}},\ }\href {\doibase
  10.1016/j.physletb.2019.134994} {\bibfield  {journal} {\bibinfo  {journal}
  {Phys. Lett.}\ }\textbf {\bibinfo {volume} {B798}},\ \bibinfo {pages}
  {134994} (\bibinfo {year} {2019})},\ \Eprint
  {http://arxiv.org/abs/1908.03331} {arXiv:1908.03331 [hep-ph]}\BibitemShut
  {NoStop}%
\bibitem [{\citenamefont {Colangelo}\ \emph
  {et~al.}(2020{\natexlab{a}})\citenamefont {Colangelo}, \citenamefont
  {Hagelstein}, \citenamefont {Hoferichter}, \citenamefont {Laub},\ and\
  \citenamefont {Stoffer}}]{Colangelo:2019lpu}%
  \BibitemOpen
  \bibfield  {author} {\bibinfo {author} {\bibfnamefont {G.}~\bibnamefont
  {Colangelo}}, \bibinfo {author} {\bibfnamefont {F.}~\bibnamefont
  {Hagelstein}}, \bibinfo {author} {\bibfnamefont {M.}~\bibnamefont
  {Hoferichter}}, \bibinfo {author} {\bibfnamefont {L.}~\bibnamefont {Laub}}, \
  and\ \bibinfo {author} {\bibfnamefont {P.}~\bibnamefont {Stoffer}},\ }\href
  {\doibase 10.1103/PhysRevD.101.051501} {\bibfield  {journal} {\bibinfo
  {journal} {Phys. Rev.}\ }\textbf {\bibinfo {volume} {D101}},\ \bibinfo
  {pages} {051501} (\bibinfo {year} {2020}{\natexlab{a}})},\ \Eprint
  {http://arxiv.org/abs/1910.11881} {arXiv:1910.11881 [hep-ph]}\BibitemShut
  {NoStop}%
\bibitem [{\citenamefont {Colangelo}\ \emph
  {et~al.}(2020{\natexlab{b}})\citenamefont {Colangelo}, \citenamefont
  {Hagelstein}, \citenamefont {Hoferichter}, \citenamefont {Laub},\ and\
  \citenamefont {Stoffer}}]{Colangelo:2019uex}%
  \BibitemOpen
  \bibfield  {author} {\bibinfo {author} {\bibfnamefont {G.}~\bibnamefont
  {Colangelo}}, \bibinfo {author} {\bibfnamefont {F.}~\bibnamefont
  {Hagelstein}}, \bibinfo {author} {\bibfnamefont {M.}~\bibnamefont
  {Hoferichter}}, \bibinfo {author} {\bibfnamefont {L.}~\bibnamefont {Laub}}, \
  and\ \bibinfo {author} {\bibfnamefont {P.}~\bibnamefont {Stoffer}},\ }\href
  {\doibase 10.1007/JHEP03(2020)101} {\bibfield  {journal} {\bibinfo  {journal}
  {JHEP}\ }\textbf {\bibinfo {volume} {03}},\ \bibinfo {pages} {101} (\bibinfo
  {year} {2020}{\natexlab{b}})},\ \Eprint {http://arxiv.org/abs/1910.13432}
  {arXiv:1910.13432 [hep-ph]}\BibitemShut {NoStop}%
\bibitem [{\citenamefont {Blum}\ \emph {et~al.}(2020)\citenamefont {Blum},
  \citenamefont {Christ}, \citenamefont {Hayakawa}, \citenamefont {Izubuchi},
  \citenamefont {Jin}, \citenamefont {Jung},\ and\ \citenamefont
  {Lehner}}]{Blum:2019ugy}%
  \BibitemOpen
  \bibfield  {author} {\bibinfo {author} {\bibfnamefont {T.}~\bibnamefont
  {Blum}}, \bibinfo {author} {\bibfnamefont {N.}~\bibnamefont {Christ}},
  \bibinfo {author} {\bibfnamefont {M.}~\bibnamefont {Hayakawa}}, \bibinfo
  {author} {\bibfnamefont {T.}~\bibnamefont {Izubuchi}}, \bibinfo {author}
  {\bibfnamefont {L.}~\bibnamefont {Jin}}, \bibinfo {author} {\bibfnamefont
  {C.}~\bibnamefont {Jung}}, \ and\ \bibinfo {author} {\bibfnamefont
  {C.}~\bibnamefont {Lehner}},\ }\href {\doibase
  10.1103/PhysRevLett.124.132002} {\bibfield  {journal} {\bibinfo  {journal}
  {Phys. Rev. Lett.}\ }\textbf {\bibinfo {volume} {124}},\ \bibinfo {pages}
  {132002} (\bibinfo {year} {2020})},\ \Eprint
  {http://arxiv.org/abs/1911.08123} {arXiv:1911.08123 [hep-lat]}\BibitemShut
  {NoStop}%
\bibitem [{\citenamefont {Colangelo}\ \emph
  {et~al.}(2014{\natexlab{c}})\citenamefont {Colangelo}, \citenamefont
  {Hoferichter}, \citenamefont {Nyffeler}, \citenamefont {Passera},\ and\
  \citenamefont {Stoffer}}]{Colangelo:2014qya}%
  \BibitemOpen
  \bibfield  {author} {\bibinfo {author} {\bibfnamefont {G.}~\bibnamefont
  {Colangelo}}, \bibinfo {author} {\bibfnamefont {M.}~\bibnamefont
  {Hoferichter}}, \bibinfo {author} {\bibfnamefont {A.}~\bibnamefont
  {Nyffeler}}, \bibinfo {author} {\bibfnamefont {M.}~\bibnamefont {Passera}}, \
  and\ \bibinfo {author} {\bibfnamefont {P.}~\bibnamefont {Stoffer}},\ }\href
  {\doibase 10.1016/j.physletb.2014.06.012} {\bibfield  {journal} {\bibinfo
  {journal} {Phys. Lett.}\ }\textbf {\bibinfo {volume} {B735}},\ \bibinfo
  {pages} {90} (\bibinfo {year} {2014}{\natexlab{c}})},\ \Eprint
  {http://arxiv.org/abs/1403.7512} {arXiv:1403.7512 [hep-ph]}\BibitemShut
  {NoStop}%
\bibitem [{\citenamefont {Bennett}\ \emph {et~al.}(2006)\citenamefont {Bennett}
  \emph {et~al.}}]{bennett:2006fi}%
  \BibitemOpen
  \bibfield  {author} {\bibinfo {author} {\bibfnamefont {G.~W.}\ \bibnamefont
  {Bennett}} \emph {et~al.} (\bibinfo {collaboration} {Muon $g-2$}),\ }\href
  {\doibase 10.1103/PhysRevD.73.072003} {\bibfield  {journal} {\bibinfo
  {journal} {Phys. Rev.}\ }\textbf {\bibinfo {volume} {D73}},\ \bibinfo {pages}
  {072003} (\bibinfo {year} {2006})},\ \Eprint
  {http://arxiv.org/abs/hep-ex/0602035} {arXiv:hep-ex/0602035
  [hep-ex]}\BibitemShut {NoStop}%
\bibitem [{\citenamefont {Chakraborty}\ \emph {et~al.}(2018)\citenamefont
  {Chakraborty} \emph {et~al.}}]{Chakraborty:2017tqp}%
  \BibitemOpen
  \bibfield  {author} {\bibinfo {author} {\bibfnamefont {B.}~\bibnamefont
  {Chakraborty}} \emph {et~al.} (\bibinfo {collaboration} {Fermilab Lattice,
  LATTICE-HPQCD, MILC}),\ }\href {\doibase 10.1103/PhysRevLett.120.152001}
  {\bibfield  {journal} {\bibinfo  {journal} {Phys. Rev. Lett.}\ }\textbf
  {\bibinfo {volume} {120}},\ \bibinfo {pages} {152001} (\bibinfo {year}
  {2018})},\ \Eprint {http://arxiv.org/abs/1710.11212} {arXiv:1710.11212
  [hep-lat]}\BibitemShut {NoStop}%
\bibitem [{\citenamefont {Borsanyi}\ \emph {et~al.}(2018)\citenamefont
  {Borsanyi} \emph {et~al.}}]{Borsanyi:2017zdw}%
  \BibitemOpen
  \bibfield  {author} {\bibinfo {author} {\bibfnamefont {S.}~\bibnamefont
  {Borsanyi}} \emph {et~al.} (\bibinfo {collaboration}
  {Budapest-Marseille-Wuppertal}),\ }\href {\doibase
  10.1103/PhysRevLett.121.022002} {\bibfield  {journal} {\bibinfo  {journal}
  {Phys. Rev. Lett.}\ }\textbf {\bibinfo {volume} {121}},\ \bibinfo {pages}
  {022002} (\bibinfo {year} {2018})},\ \Eprint
  {http://arxiv.org/abs/1711.04980} {arXiv:1711.04980 [hep-lat]}\BibitemShut
  {NoStop}%
\bibitem [{\citenamefont {Blum}\ \emph {et~al.}(2018)\citenamefont {Blum},
  \citenamefont {Boyle}, \citenamefont {G{\"u}lpers}, \citenamefont {Izubuchi},
  \citenamefont {Jin}, \citenamefont {Jung}, \citenamefont {J{\"u}ttner},
  \citenamefont {Lehner}, \citenamefont {Portelli},\ and\ \citenamefont
  {Tsang}}]{Blum:2018mom}%
  \BibitemOpen
  \bibfield  {author} {\bibinfo {author} {\bibfnamefont {T.}~\bibnamefont
  {Blum}}, \bibinfo {author} {\bibfnamefont {P.~A.}\ \bibnamefont {Boyle}},
  \bibinfo {author} {\bibfnamefont {V.}~\bibnamefont {G{\"u}lpers}}, \bibinfo
  {author} {\bibfnamefont {T.}~\bibnamefont {Izubuchi}}, \bibinfo {author}
  {\bibfnamefont {L.}~\bibnamefont {Jin}}, \bibinfo {author} {\bibfnamefont
  {C.}~\bibnamefont {Jung}}, \bibinfo {author} {\bibfnamefont {A.}~\bibnamefont
  {J{\"u}ttner}}, \bibinfo {author} {\bibfnamefont {C.}~\bibnamefont {Lehner}},
  \bibinfo {author} {\bibfnamefont {A.}~\bibnamefont {Portelli}}, \ and\
  \bibinfo {author} {\bibfnamefont {J.~T.}\ \bibnamefont {Tsang}} (\bibinfo
  {collaboration} {RBC, UKQCD}),\ }\href {\doibase
  10.1103/PhysRevLett.121.022003} {\bibfield  {journal} {\bibinfo  {journal}
  {Phys. Rev. Lett.}\ }\textbf {\bibinfo {volume} {121}},\ \bibinfo {pages}
  {022003} (\bibinfo {year} {2018})},\ \Eprint
  {http://arxiv.org/abs/1801.07224} {arXiv:1801.07224 [hep-lat]}\BibitemShut
  {NoStop}%
\bibitem [{\citenamefont {Giusti}\ \emph {et~al.}(2019)\citenamefont {Giusti},
  \citenamefont {Lubicz}, \citenamefont {Martinelli}, \citenamefont
  {Sanfilippo},\ and\ \citenamefont {Simula}}]{Giusti:2019xct}%
  \BibitemOpen
  \bibfield  {author} {\bibinfo {author} {\bibfnamefont {D.}~\bibnamefont
  {Giusti}}, \bibinfo {author} {\bibfnamefont {V.}~\bibnamefont {Lubicz}},
  \bibinfo {author} {\bibfnamefont {G.}~\bibnamefont {Martinelli}}, \bibinfo
  {author} {\bibfnamefont {F.}~\bibnamefont {Sanfilippo}}, \ and\ \bibinfo
  {author} {\bibfnamefont {S.}~\bibnamefont {Simula}} (\bibinfo {collaboration}
  {ETM}),\ }\href {\doibase 10.1103/PhysRevD.99.114502} {\bibfield  {journal}
  {\bibinfo  {journal} {Phys. Rev.}\ }\textbf {\bibinfo {volume} {D99}},\
  \bibinfo {pages} {114502} (\bibinfo {year} {2019})},\ \Eprint
  {http://arxiv.org/abs/1901.10462} {arXiv:1901.10462 [hep-lat]}\BibitemShut
  {NoStop}%
\bibitem [{\citenamefont {Shintani}\ and\ \citenamefont
  {Kuramashi}(2019)}]{Shintani:2019wai}%
  \BibitemOpen
  \bibfield  {author} {\bibinfo {author} {\bibfnamefont {E.}~\bibnamefont
  {Shintani}}\ and\ \bibinfo {author} {\bibfnamefont {Y.}~\bibnamefont
  {Kuramashi}},\ }\href {\doibase 10.1103/PhysRevD.100.034517} {\bibfield
  {journal} {\bibinfo  {journal} {Phys. Rev.}\ }\textbf {\bibinfo {volume}
  {D100}},\ \bibinfo {pages} {034517} (\bibinfo {year} {2019})},\ \Eprint
  {http://arxiv.org/abs/1902.00885} {arXiv:1902.00885 [hep-lat]}\BibitemShut
  {NoStop}%
\bibitem [{\citenamefont {Davies}\ \emph {et~al.}(2020)\citenamefont {Davies}
  \emph {et~al.}}]{Davies:2019efs}%
  \BibitemOpen
  \bibfield  {author} {\bibinfo {author} {\bibfnamefont {C.~T.~H.}\
  \bibnamefont {Davies}} \emph {et~al.} (\bibinfo {collaboration} {Fermilab
  Lattice, LATTICE-HPQCD, MILC}),\ }\href {\doibase
  10.1103/PhysRevD.101.034512} {\bibfield  {journal} {\bibinfo  {journal}
  {Phys. Rev.}\ }\textbf {\bibinfo {volume} {D101}},\ \bibinfo {pages} {034512}
  (\bibinfo {year} {2020})},\ \Eprint {http://arxiv.org/abs/1902.04223}
  {arXiv:1902.04223 [hep-lat]}\BibitemShut {NoStop}%
\bibitem [{\citenamefont {G\'erardin}\ \emph {et~al.}(2019)\citenamefont
  {G\'erardin}, \citenamefont {C\`e}, \citenamefont {von Hippel}, \citenamefont
  {H{\"o}rz}, \citenamefont {Meyer}, \citenamefont {Mohler}, \citenamefont
  {Ottnad}, \citenamefont {Wilhelm},\ and\ \citenamefont
  {Wittig}}]{Gerardin:2019rua}%
  \BibitemOpen
  \bibfield  {author} {\bibinfo {author} {\bibfnamefont {A.}~\bibnamefont
  {G\'erardin}}, \bibinfo {author} {\bibfnamefont {M.}~\bibnamefont {C\`e}},
  \bibinfo {author} {\bibfnamefont {G.}~\bibnamefont {von Hippel}}, \bibinfo
  {author} {\bibfnamefont {B.}~\bibnamefont {H{\"o}rz}}, \bibinfo {author}
  {\bibfnamefont {H.~B.}\ \bibnamefont {Meyer}}, \bibinfo {author}
  {\bibfnamefont {D.}~\bibnamefont {Mohler}}, \bibinfo {author} {\bibfnamefont
  {K.}~\bibnamefont {Ottnad}}, \bibinfo {author} {\bibfnamefont
  {J.}~\bibnamefont {Wilhelm}}, \ and\ \bibinfo {author} {\bibfnamefont
  {H.}~\bibnamefont {Wittig}},\ }\href {\doibase 10.1103/PhysRevD.100.014510}
  {\bibfield  {journal} {\bibinfo  {journal} {Phys. Rev.}\ }\textbf {\bibinfo
  {volume} {D100}},\ \bibinfo {pages} {014510} (\bibinfo {year} {2019})},\
  \Eprint {http://arxiv.org/abs/1904.03120} {arXiv:1904.03120
  [hep-lat]}\BibitemShut {NoStop}%
\bibitem [{\citenamefont {Aubin}\ \emph {et~al.}(2020)\citenamefont {Aubin},
  \citenamefont {Blum}, \citenamefont {Tu}, \citenamefont {Golterman},
  \citenamefont {Jung},\ and\ \citenamefont {Peris}}]{Aubin:2019usy}%
  \BibitemOpen
  \bibfield  {author} {\bibinfo {author} {\bibfnamefont {C.}~\bibnamefont
  {Aubin}}, \bibinfo {author} {\bibfnamefont {T.}~\bibnamefont {Blum}},
  \bibinfo {author} {\bibfnamefont {C.}~\bibnamefont {Tu}}, \bibinfo {author}
  {\bibfnamefont {M.}~\bibnamefont {Golterman}}, \bibinfo {author}
  {\bibfnamefont {C.}~\bibnamefont {Jung}}, \ and\ \bibinfo {author}
  {\bibfnamefont {S.}~\bibnamefont {Peris}},\ }\href {\doibase
  10.1103/PhysRevD.101.014503} {\bibfield  {journal} {\bibinfo  {journal}
  {Phys. Rev.}\ }\textbf {\bibinfo {volume} {D101}},\ \bibinfo {pages} {014503}
  (\bibinfo {year} {2020})},\ \Eprint {http://arxiv.org/abs/1905.09307}
  {arXiv:1905.09307 [hep-lat]}\BibitemShut {NoStop}%
\bibitem [{\citenamefont {Giusti}\ and\ \citenamefont
  {Simula}(2019)}]{Giusti:2019hkz}%
  \BibitemOpen
  \bibfield  {author} {\bibinfo {author} {\bibfnamefont {D.}~\bibnamefont
  {Giusti}}\ and\ \bibinfo {author} {\bibfnamefont {S.}~\bibnamefont
  {Simula}},\ }\href {\doibase 10.22323/1.363.0104} {\bibfield  {journal}
  {\bibinfo  {journal} {PoS}\ }\textbf {\bibinfo {volume} {LATTICE2019}},\
  \bibinfo {pages} {104} (\bibinfo {year} {2019})},\ \Eprint
  {http://arxiv.org/abs/1910.03874} {arXiv:1910.03874 [hep-lat]}\BibitemShut
  {NoStop}%
\bibitem [{\citenamefont {Borsanyi}\ \emph {et~al.}(2020)\citenamefont
  {Borsanyi} \emph {et~al.}}]{Borsanyi:2020mff}%
  \BibitemOpen
  \bibfield  {author} {\bibinfo {author} {\bibfnamefont {S.}~\bibnamefont
  {Borsanyi}} \emph {et~al.},\ }\href@noop {} {\  (\bibinfo {year} {2020})},\
  \Eprint {http://arxiv.org/abs/2002.12347} {arXiv:2002.12347
  [hep-lat]}\BibitemShut {NoStop}%
\bibitem [{\citenamefont {Pauk}\ and\ \citenamefont
  {Vanderhaeghen}(2014)}]{Pauk:2014rta}%
  \BibitemOpen
  \bibfield  {author} {\bibinfo {author} {\bibfnamefont {V.}~\bibnamefont
  {Pauk}}\ and\ \bibinfo {author} {\bibfnamefont {M.}~\bibnamefont
  {Vanderhaeghen}},\ }\href {\doibase 10.1140/epjc/s10052-014-3008-y}
  {\bibfield  {journal} {\bibinfo  {journal} {Eur. Phys. J.}\ }\textbf
  {\bibinfo {volume} {C74}},\ \bibinfo {pages} {3008} (\bibinfo {year}
  {2014})},\ \Eprint {http://arxiv.org/abs/1401.0832} {arXiv:1401.0832
  [hep-ph]}\BibitemShut {NoStop}%
\bibitem [{\citenamefont {Danilkin}\ and\ \citenamefont
  {Vanderhaeghen}(2017)}]{Danilkin:2016hnh}%
  \BibitemOpen
  \bibfield  {author} {\bibinfo {author} {\bibfnamefont {I.}~\bibnamefont
  {Danilkin}}\ and\ \bibinfo {author} {\bibfnamefont {M.}~\bibnamefont
  {Vanderhaeghen}},\ }\href {\doibase 10.1103/PhysRevD.95.014019} {\bibfield
  {journal} {\bibinfo  {journal} {Phys. Rev.}\ }\textbf {\bibinfo {volume}
  {D95}},\ \bibinfo {pages} {014019} (\bibinfo {year} {2017})},\ \Eprint
  {http://arxiv.org/abs/1611.04646} {arXiv:1611.04646 [hep-ph]}\BibitemShut
  {NoStop}%
\bibitem [{\citenamefont {Jegerlehner}(2017)}]{Jegerlehner:2017gek}%
  \BibitemOpen
  \bibfield  {author} {\bibinfo {author} {\bibfnamefont {F.}~\bibnamefont
  {Jegerlehner}},\ }\href {\doibase 10.1007/978-3-319-63577-4} {\bibfield
  {journal} {\bibinfo  {journal} {Springer Tracts Mod. Phys.}\ }\textbf
  {\bibinfo {volume} {274}},\ \bibinfo {pages} {1} (\bibinfo {year}
  {2017})}\BibitemShut {NoStop}%
\bibitem [{\citenamefont {Knecht}\ \emph {et~al.}(2018)\citenamefont {Knecht},
  \citenamefont {Narison}, \citenamefont {Rabemananjara},\ and\ \citenamefont
  {Rabetiarivony}}]{Knecht:2018sci}%
  \BibitemOpen
  \bibfield  {author} {\bibinfo {author} {\bibfnamefont {M.}~\bibnamefont
  {Knecht}}, \bibinfo {author} {\bibfnamefont {S.}~\bibnamefont {Narison}},
  \bibinfo {author} {\bibfnamefont {A.}~\bibnamefont {Rabemananjara}}, \ and\
  \bibinfo {author} {\bibfnamefont {D.}~\bibnamefont {Rabetiarivony}},\ }\href
  {\doibase 10.1016/j.physletb.2018.10.048} {\bibfield  {journal} {\bibinfo
  {journal} {Phys. Lett.}\ }\textbf {\bibinfo {volume} {B787}},\ \bibinfo
  {pages} {111} (\bibinfo {year} {2018})},\ \Eprint
  {http://arxiv.org/abs/1808.03848} {arXiv:1808.03848 [hep-ph]}\BibitemShut
  {NoStop}%
\bibitem [{\citenamefont {Eichmann}\ \emph {et~al.}(2020)\citenamefont
  {Eichmann}, \citenamefont {Fischer},\ and\ \citenamefont
  {Williams}}]{Eichmann:2019bqf}%
  \BibitemOpen
  \bibfield  {author} {\bibinfo {author} {\bibfnamefont {G.}~\bibnamefont
  {Eichmann}}, \bibinfo {author} {\bibfnamefont {C.~S.}\ \bibnamefont
  {Fischer}}, \ and\ \bibinfo {author} {\bibfnamefont {R.}~\bibnamefont
  {Williams}},\ }\href {\doibase 10.1103/PhysRevD.101.054015} {\bibfield
  {journal} {\bibinfo  {journal} {Phys. Rev.}\ }\textbf {\bibinfo {volume}
  {D101}},\ \bibinfo {pages} {054015} (\bibinfo {year} {2020})},\ \Eprint
  {http://arxiv.org/abs/1910.06795} {arXiv:1910.06795 [hep-ph]}\BibitemShut
  {NoStop}%
\bibitem [{\citenamefont {Roig}\ and\ \citenamefont
  {S{\'a}nchez-Puertas}(2020)}]{Roig:2019reh}%
  \BibitemOpen
  \bibfield  {author} {\bibinfo {author} {\bibfnamefont {P.}~\bibnamefont
  {Roig}}\ and\ \bibinfo {author} {\bibfnamefont {P.}~\bibnamefont
  {S{\'a}nchez-Puertas}},\ }\href {\doibase 10.1103/PhysRevD.101.074019}
  {\bibfield  {journal} {\bibinfo  {journal} {Phys. Rev.}\ }\textbf {\bibinfo
  {volume} {D101}},\ \bibinfo {pages} {074019} (\bibinfo {year} {2020})},\
  \Eprint {http://arxiv.org/abs/1910.02881} {arXiv:1910.02881
  [hep-ph]}\BibitemShut {NoStop}%
\bibitem [{\citenamefont {Passera}\ \emph {et~al.}(2008)\citenamefont
  {Passera}, \citenamefont {Marciano},\ and\ \citenamefont
  {Sirlin}}]{Passera:2008jk}%
  \BibitemOpen
  \bibfield  {author} {\bibinfo {author} {\bibfnamefont {M.}~\bibnamefont
  {Passera}}, \bibinfo {author} {\bibfnamefont {W.}~\bibnamefont {Marciano}}, \
  and\ \bibinfo {author} {\bibfnamefont {A.}~\bibnamefont {Sirlin}},\ }\href
  {\doibase 10.1103/PhysRevD.78.013009} {\bibfield  {journal} {\bibinfo
  {journal} {Phys. Rev.}\ }\textbf {\bibinfo {volume} {D78}},\ \bibinfo {pages}
  {013009} (\bibinfo {year} {2008})},\ \Eprint {http://arxiv.org/abs/0804.1142}
  {arXiv:0804.1142 [hep-ph]}\BibitemShut {NoStop}%
\bibitem [{\citenamefont {Crivellin}\ \emph {et~al.}(2020)\citenamefont
  {Crivellin}, \citenamefont {Hoferichter}, \citenamefont {Manzari},\ and\
  \citenamefont {Montull}}]{Crivellin:2020zul}%
  \BibitemOpen
  \bibfield  {author} {\bibinfo {author} {\bibfnamefont {A.}~\bibnamefont
  {Crivellin}}, \bibinfo {author} {\bibfnamefont {M.}~\bibnamefont
  {Hoferichter}}, \bibinfo {author} {\bibfnamefont {C.~A.}\ \bibnamefont
  {Manzari}}, \ and\ \bibinfo {author} {\bibfnamefont {M.}~\bibnamefont
  {Montull}},\ }\href {\doibase 10.1103/PhysRevLett.125.091801} {\bibfield
  {journal} {\bibinfo  {journal} {Phys. Rev. Lett.}\ }\textbf {\bibinfo
  {volume} {125}},\ \bibinfo {pages} {091801} (\bibinfo {year} {2020})},\
  \Eprint {http://arxiv.org/abs/2003.04886} {arXiv:2003.04886
  [hep-ph]}\BibitemShut {NoStop}%
\bibitem [{\citenamefont {Keshavarzi}\ \emph
  {et~al.}(2020{\natexlab{b}})\citenamefont {Keshavarzi}, \citenamefont
  {Marciano}, \citenamefont {Passera},\ and\ \citenamefont
  {Sirlin}}]{Keshavarzi:2020bfy}%
  \BibitemOpen
  \bibfield  {author} {\bibinfo {author} {\bibfnamefont {A.}~\bibnamefont
  {Keshavarzi}}, \bibinfo {author} {\bibfnamefont {W.~J.}\ \bibnamefont
  {Marciano}}, \bibinfo {author} {\bibfnamefont {M.}~\bibnamefont {Passera}}, \
  and\ \bibinfo {author} {\bibfnamefont {A.}~\bibnamefont {Sirlin}},\ }\href
  {\doibase 10.1103/PhysRevD.102.033002} {\bibfield  {journal} {\bibinfo
  {journal} {Phys. Rev.}\ }\textbf {\bibinfo {volume} {D102}},\ \bibinfo
  {pages} {033002} (\bibinfo {year} {2020}{\natexlab{b}})},\ \Eprint
  {http://arxiv.org/abs/2006.12666} {arXiv:2006.12666 [hep-ph]}\BibitemShut
  {NoStop}%
\bibitem [{\citenamefont {Malaescu}\ and\ \citenamefont
  {Schott}(2020)}]{Malaescu:2020zuc}%
  \BibitemOpen
  \bibfield  {author} {\bibinfo {author} {\bibfnamefont {B.}~\bibnamefont
  {Malaescu}}\ and\ \bibinfo {author} {\bibfnamefont {M.}~\bibnamefont
  {Schott}},\ }\href@noop {} {\  (\bibinfo {year} {2020})},\ \Eprint
  {http://arxiv.org/abs/2008.08107} {arXiv:2008.08107 [hep-ph]}\BibitemShut
  {NoStop}%
\bibitem [{\citenamefont {Roy}(1971)}]{Roy:1971tc}%
  \BibitemOpen
  \bibfield  {author} {\bibinfo {author} {\bibfnamefont {S.~M.}\ \bibnamefont
  {Roy}},\ }\href {\doibase 10.1016/0370-2693(71)90724-6} {\bibfield  {journal}
  {\bibinfo  {journal} {Phys. Lett.}\ }\textbf {\bibinfo {volume} {36B}},\
  \bibinfo {pages} {353} (\bibinfo {year} {1971})}\BibitemShut {NoStop}%
\bibitem [{\citenamefont {Ananthanarayan}\ \emph {et~al.}(2001)\citenamefont
  {Ananthanarayan}, \citenamefont {Colangelo}, \citenamefont {Gasser},\ and\
  \citenamefont {Leutwyler}}]{Ananthanarayan:2000ht}%
  \BibitemOpen
  \bibfield  {author} {\bibinfo {author} {\bibfnamefont {B.}~\bibnamefont
  {Ananthanarayan}}, \bibinfo {author} {\bibfnamefont {G.}~\bibnamefont
  {Colangelo}}, \bibinfo {author} {\bibfnamefont {J.}~\bibnamefont {Gasser}}, \
  and\ \bibinfo {author} {\bibfnamefont {H.}~\bibnamefont {Leutwyler}},\ }\href
  {\doibase 10.1016/S0370-1573(01)00009-6} {\bibfield  {journal} {\bibinfo
  {journal} {Phys. Rept.}\ }\textbf {\bibinfo {volume} {353}},\ \bibinfo
  {pages} {207} (\bibinfo {year} {2001})},\ \Eprint
  {http://arxiv.org/abs/hep-ph/0005297} {arXiv:hep-ph/0005297
  [hep-ph]}\BibitemShut {NoStop}%
\bibitem [{\citenamefont {Garc{\'i}a-Mart{\'i}n}\ \emph
  {et~al.}(2011)\citenamefont {Garc{\'i}a-Mart{\'i}n}, \citenamefont
  {Kami{\'n}ski}, \citenamefont {Pel{\'a}ez}, \citenamefont {Ruiz~de Elvira},\
  and\ \citenamefont {Yndur{\'a}in}}]{GarciaMartin:2011cn}%
  \BibitemOpen
  \bibfield  {author} {\bibinfo {author} {\bibfnamefont {R.}~\bibnamefont
  {Garc{\'i}a-Mart{\'i}n}}, \bibinfo {author} {\bibfnamefont {R.}~\bibnamefont
  {Kami{\'n}ski}}, \bibinfo {author} {\bibfnamefont {J.~R.}\ \bibnamefont
  {Pel{\'a}ez}}, \bibinfo {author} {\bibfnamefont {J.}~\bibnamefont {Ruiz~de
  Elvira}}, \ and\ \bibinfo {author} {\bibfnamefont {F.~J.}\ \bibnamefont
  {Yndur{\'a}in}},\ }\href {\doibase 10.1103/PhysRevD.83.074004} {\bibfield
  {journal} {\bibinfo  {journal} {Phys. Rev.}\ }\textbf {\bibinfo {volume}
  {D83}},\ \bibinfo {pages} {074004} (\bibinfo {year} {2011})},\ \Eprint
  {http://arxiv.org/abs/1102.2183} {arXiv:1102.2183 [hep-ph]}\BibitemShut
  {NoStop}%
\bibitem [{\citenamefont {Caprini}\ \emph {et~al.}(2012)\citenamefont
  {Caprini}, \citenamefont {Colangelo},\ and\ \citenamefont
  {Leutwyler}}]{Caprini:2011ky}%
  \BibitemOpen
  \bibfield  {author} {\bibinfo {author} {\bibfnamefont {I.}~\bibnamefont
  {Caprini}}, \bibinfo {author} {\bibfnamefont {G.}~\bibnamefont {Colangelo}},
  \ and\ \bibinfo {author} {\bibfnamefont {H.}~\bibnamefont {Leutwyler}},\
  }\href {\doibase 10.1140/epjc/s10052-012-1860-1} {\bibfield  {journal}
  {\bibinfo  {journal} {Eur. Phys. J.}\ }\textbf {\bibinfo {volume} {C72}},\
  \bibinfo {pages} {1860} (\bibinfo {year} {2012})},\ \Eprint
  {http://arxiv.org/abs/1111.7160} {arXiv:1111.7160 [hep-ph]}\BibitemShut
  {NoStop}%
\bibitem [{\citenamefont {Bouchiat}\ and\ \citenamefont
  {Michel}(1961)}]{Bouchiat:1961lbg}%
  \BibitemOpen
  \bibfield  {author} {\bibinfo {author} {\bibfnamefont {C.}~\bibnamefont
  {Bouchiat}}\ and\ \bibinfo {author} {\bibfnamefont {L.}~\bibnamefont
  {Michel}},\ }\href {\doibase 10.1051/jphysrad:01961002202012101} {\bibfield
  {journal} {\bibinfo  {journal} {J. Phys. Radium}\ }\textbf {\bibinfo {volume}
  {22}},\ \bibinfo {pages} {121} (\bibinfo {year} {1961})}\BibitemShut
  {NoStop}%
\bibitem [{\citenamefont {Brodsky}\ and\ \citenamefont
  {de~Rafael}(1968)}]{Brodsky:1967sr}%
  \BibitemOpen
  \bibfield  {author} {\bibinfo {author} {\bibfnamefont {S.~J.}\ \bibnamefont
  {Brodsky}}\ and\ \bibinfo {author} {\bibfnamefont {E.}~\bibnamefont
  {de~Rafael}},\ }\href {\doibase 10.1103/PhysRev.168.1620} {\bibfield
  {journal} {\bibinfo  {journal} {Phys. Rev.}\ }\textbf {\bibinfo {volume}
  {168}},\ \bibinfo {pages} {1620} (\bibinfo {year} {1968})}\BibitemShut
  {NoStop}%
\bibitem [{\citenamefont {Omn{\`e}s}(1958)}]{Omnes:1958hv}%
  \BibitemOpen
  \bibfield  {author} {\bibinfo {author} {\bibfnamefont {R.}~\bibnamefont
  {Omn{\`e}s}},\ }\href {\doibase 10.1007/BF02747746} {\bibfield  {journal}
  {\bibinfo  {journal} {Nuovo Cim.}\ }\textbf {\bibinfo {volume} {8}},\
  \bibinfo {pages} {316} (\bibinfo {year} {1958})}\BibitemShut {NoStop}%
\bibitem [{\citenamefont {de~Troc{\'o}niz}\ and\ \citenamefont
  {Yndur{\'a}in}(2002)}]{DeTroconiz:2001rip}%
  \BibitemOpen
  \bibfield  {author} {\bibinfo {author} {\bibfnamefont {J.~F.}\ \bibnamefont
  {de~Troc{\'o}niz}}\ and\ \bibinfo {author} {\bibfnamefont {F.~J.}\
  \bibnamefont {Yndur{\'a}in}},\ }\href {\doibase 10.1103/PhysRevD.65.093001}
  {\bibfield  {journal} {\bibinfo  {journal} {Phys. Rev.}\ }\textbf {\bibinfo
  {volume} {D65}},\ \bibinfo {pages} {093001} (\bibinfo {year} {2002})},\
  \Eprint {http://arxiv.org/abs/hep-ph/0106025} {arXiv:hep-ph/0106025
  [hep-ph]}\BibitemShut {NoStop}%
\bibitem [{\citenamefont {Leutwyler}(2002)}]{Leutwyler:2002hm}%
  \BibitemOpen
  \bibfield  {author} {\bibinfo {author} {\bibfnamefont {H.}~\bibnamefont
  {Leutwyler}},\ }\href {\doibase 10.1142/9789812776310_0002} {\bibfield
  {journal} {\bibinfo  {journal} {{Continuous advances in QCD}}\ }\textbf
  {\bibinfo {volume} {2002}},\ \bibinfo {pages} {23} (\bibinfo {year}
  {2002})},\ \Eprint {http://arxiv.org/abs/hep-ph/0212324}
  {arXiv:hep-ph/0212324 [hep-ph]}\BibitemShut {NoStop}%
\bibitem [{\citenamefont {Colangelo}(2004)}]{Colangelo:2003yw}%
  \BibitemOpen
  \bibfield  {author} {\bibinfo {author} {\bibfnamefont {G.}~\bibnamefont
  {Colangelo}},\ }\href {\doibase 10.1016/j.nuclphysbps.2004.02.025} {\bibfield
   {journal} {\bibinfo  {journal} {Nucl. Phys. Proc. Suppl.}\ }\textbf
  {\bibinfo {volume} {131}},\ \bibinfo {pages} {185} (\bibinfo {year}
  {2004})},\ \Eprint {http://arxiv.org/abs/hep-ph/0312017}
  {arXiv:hep-ph/0312017 [hep-ph]}\BibitemShut {NoStop}%
\bibitem [{\citenamefont {de~Troc{\'o}niz}\ and\ \citenamefont
  {Yndur{\'a}in}(2005)}]{deTroconiz:2004yzs}%
  \BibitemOpen
  \bibfield  {author} {\bibinfo {author} {\bibfnamefont {J.~F.}\ \bibnamefont
  {de~Troc{\'o}niz}}\ and\ \bibinfo {author} {\bibfnamefont {F.~J.}\
  \bibnamefont {Yndur{\'a}in}},\ }\href {\doibase 10.1103/PhysRevD.71.073008}
  {\bibfield  {journal} {\bibinfo  {journal} {Phys. Rev.}\ }\textbf {\bibinfo
  {volume} {D71}},\ \bibinfo {pages} {073008} (\bibinfo {year} {2005})},\
  \Eprint {http://arxiv.org/abs/hep-ph/0402285} {arXiv:hep-ph/0402285
  [hep-ph]}\BibitemShut {NoStop}%
\bibitem [{\citenamefont {Hoferichter}\ \emph {et~al.}(2016)\citenamefont
  {Hoferichter}, \citenamefont {Kubis}, \citenamefont {Ruiz~de Elvira},
  \citenamefont {Hammer},\ and\ \citenamefont
  {Mei{\ss}ner}}]{Hoferichter:2016duk}%
  \BibitemOpen
  \bibfield  {author} {\bibinfo {author} {\bibfnamefont {M.}~\bibnamefont
  {Hoferichter}}, \bibinfo {author} {\bibfnamefont {B.}~\bibnamefont {Kubis}},
  \bibinfo {author} {\bibfnamefont {J.}~\bibnamefont {Ruiz~de Elvira}},
  \bibinfo {author} {\bibfnamefont {H.-W.}\ \bibnamefont {Hammer}}, \ and\
  \bibinfo {author} {\bibfnamefont {U.-G.}\ \bibnamefont {Mei{\ss}ner}},\
  }\href {\doibase 10.1140/epja/i2016-16331-7} {\bibfield  {journal} {\bibinfo
  {journal} {Eur. Phys. J.}\ }\textbf {\bibinfo {volume} {A52}},\ \bibinfo
  {pages} {331} (\bibinfo {year} {2016})},\ \Eprint
  {http://arxiv.org/abs/1609.06722} {arXiv:1609.06722 [hep-ph]}\BibitemShut
  {NoStop}%
\bibitem [{\citenamefont {Hanhart}\ \emph {et~al.}(2017)\citenamefont
  {Hanhart}, \citenamefont {Holz}, \citenamefont {Kubis}, \citenamefont
  {Kup{\'s}{\'c}}, \citenamefont {Wirzba},\ and\ \citenamefont
  {Xiao}}]{Hanhart:2016pcd}%
  \BibitemOpen
  \bibfield  {author} {\bibinfo {author} {\bibfnamefont {C.}~\bibnamefont
  {Hanhart}}, \bibinfo {author} {\bibfnamefont {S.}~\bibnamefont {Holz}},
  \bibinfo {author} {\bibfnamefont {B.}~\bibnamefont {Kubis}}, \bibinfo
  {author} {\bibfnamefont {A.}~\bibnamefont {Kup{\'s}{\'c}}}, \bibinfo {author}
  {\bibfnamefont {A.}~\bibnamefont {Wirzba}}, \ and\ \bibinfo {author}
  {\bibfnamefont {C.~W.}\ \bibnamefont {Xiao}},\ }\href {\doibase
  10.1140/epjc/s10052-018-5941-7, 10.1140/epjc/s10052-017-4651-x} {\bibfield
  {journal} {\bibinfo  {journal} {Eur. Phys. J.}\ }\textbf {\bibinfo {volume}
  {C77}},\ \bibinfo {pages} {98} (\bibinfo {year} {2017})},\ \bibinfo {note}
  {[Erratum: Eur. Phys. J. {\bf C78}, 450 (2018)]},\ \Eprint
  {http://arxiv.org/abs/1611.09359} {arXiv:1611.09359 [hep-ph]}\BibitemShut
  {NoStop}%
\bibitem [{\citenamefont {Ananthanarayan}\ \emph {et~al.}(2018)\citenamefont
  {Ananthanarayan}, \citenamefont {Caprini},\ and\ \citenamefont
  {Das}}]{Ananthanarayan:2018nyx}%
  \BibitemOpen
  \bibfield  {author} {\bibinfo {author} {\bibfnamefont {B.}~\bibnamefont
  {Ananthanarayan}}, \bibinfo {author} {\bibfnamefont {I.}~\bibnamefont
  {Caprini}}, \ and\ \bibinfo {author} {\bibfnamefont {D.}~\bibnamefont
  {Das}},\ }\href {\doibase 10.1103/PhysRevD.98.114015} {\bibfield  {journal}
  {\bibinfo  {journal} {Phys. Rev.}\ }\textbf {\bibinfo {volume} {D98}},\
  \bibinfo {pages} {114015} (\bibinfo {year} {2018})},\ \Eprint
  {http://arxiv.org/abs/1810.09265} {arXiv:1810.09265 [hep-ph]}\BibitemShut
  {NoStop}%
\bibitem [{\citenamefont {Ananthanarayan}\ \emph {et~al.}(2020)\citenamefont
  {Ananthanarayan}, \citenamefont {Caprini},\ and\ \citenamefont
  {Das}}]{Ananthanarayan:2020vum}%
  \BibitemOpen
  \bibfield  {author} {\bibinfo {author} {\bibfnamefont {B.}~\bibnamefont
  {Ananthanarayan}}, \bibinfo {author} {\bibfnamefont {I.}~\bibnamefont
  {Caprini}}, \ and\ \bibinfo {author} {\bibfnamefont {D.}~\bibnamefont
  {Das}},\ }\href {\doibase 10.1103/PhysRevD.102.096003} {\bibfield  {journal}
  {\bibinfo  {journal} {Phys. Rev. D}\ }\textbf {\bibinfo {volume} {102}},\
  \bibinfo {pages} {096003} (\bibinfo {year} {2020})},\ \Eprint
  {http://arxiv.org/abs/2008.00669} {arXiv:2008.00669 [hep-ph]}\BibitemShut
  {NoStop}%
\bibitem [{\citenamefont {Batley}\ \emph {et~al.}(2010)\citenamefont {Batley}
  \emph {et~al.}}]{Batley:2010zza}%
  \BibitemOpen
  \bibfield  {author} {\bibinfo {author} {\bibfnamefont {J.}~\bibnamefont
  {Batley}} \emph {et~al.} (\bibinfo {collaboration} {NA48/2}),\ }\href
  {\doibase 10.1140/epjc/s10052-010-1480-6} {\bibfield  {journal} {\bibinfo
  {journal} {Eur. Phys. J.}\ }\textbf {\bibinfo {volume} {C70}},\ \bibinfo
  {pages} {635} (\bibinfo {year} {2010})}\BibitemShut {NoStop}%
\bibitem [{\citenamefont {Batley}\ \emph {et~al.}(2012)\citenamefont {Batley}
  \emph {et~al.}}]{Batley:2012rf}%
  \BibitemOpen
  \bibfield  {author} {\bibinfo {author} {\bibfnamefont {J.}~\bibnamefont
  {Batley}} \emph {et~al.} (\bibinfo {collaboration} {NA48/2}),\ }\href
  {\doibase 10.1016/j.physletb.2012.07.048} {\bibfield  {journal} {\bibinfo
  {journal} {Phys. Lett.}\ }\textbf {\bibinfo {volume} {B715}},\ \bibinfo
  {pages} {105} (\bibinfo {year} {2012})},\ \bibinfo {note} {[Addendum: Phys.
  Lett. {\bf B740}, 364 (2015)]},\ \Eprint {http://arxiv.org/abs/1206.7065}
  {arXiv:1206.7065 [hep-ex]}\BibitemShut {NoStop}%
\bibitem [{\citenamefont {Colangelo}\ \emph
  {et~al.}(2015{\natexlab{b}})\citenamefont {Colangelo}, \citenamefont
  {Passemar},\ and\ \citenamefont {Stoffer}}]{Colangelo:2015kha}%
  \BibitemOpen
  \bibfield  {author} {\bibinfo {author} {\bibfnamefont {G.}~\bibnamefont
  {Colangelo}}, \bibinfo {author} {\bibfnamefont {E.}~\bibnamefont {Passemar}},
  \ and\ \bibinfo {author} {\bibfnamefont {P.}~\bibnamefont {Stoffer}},\ }\href
  {\doibase 10.1140/epjc/s10052-015-3357-1} {\bibfield  {journal} {\bibinfo
  {journal} {Eur. Phys. J.}\ }\textbf {\bibinfo {volume} {C75}},\ \bibinfo
  {pages} {172} (\bibinfo {year} {2015}{\natexlab{b}})},\ \Eprint
  {http://arxiv.org/abs/1501.05627} {arXiv:1501.05627 [hep-ph]}\BibitemShut
  {NoStop}%
\bibitem [{\citenamefont {Hyams}\ \emph {et~al.}(1973)\citenamefont {Hyams}
  \emph {et~al.}}]{Hyams:1973zf}%
  \BibitemOpen
  \bibfield  {author} {\bibinfo {author} {\bibfnamefont {B.}~\bibnamefont
  {Hyams}} \emph {et~al.},\ }\href {\doibase 10.1016/0550-3213(73)90618-4}
  {\bibfield  {journal} {\bibinfo  {journal} {Nucl. Phys.}\ }\textbf {\bibinfo
  {volume} {B64}},\ \bibinfo {pages} {134} (\bibinfo {year}
  {1973})}\BibitemShut {NoStop}%
\bibitem [{\citenamefont {Protopopescu}\ \emph {et~al.}(1973)\citenamefont
  {Protopopescu} \emph {et~al.}}]{Protopopescu:1973sh}%
  \BibitemOpen
  \bibfield  {author} {\bibinfo {author} {\bibfnamefont {S.}~\bibnamefont
  {Protopopescu}} \emph {et~al.},\ }\href {\doibase 10.1103/PhysRevD.7.1279}
  {\bibfield  {journal} {\bibinfo  {journal} {Phys. Rev.}\ }\textbf {\bibinfo
  {volume} {D7}},\ \bibinfo {pages} {1279} (\bibinfo {year}
  {1973})}\BibitemShut {NoStop}%
\bibitem [{\citenamefont {Watson}(1954)}]{Watson:1954uc}%
  \BibitemOpen
  \bibfield  {author} {\bibinfo {author} {\bibfnamefont {K.~M.}\ \bibnamefont
  {Watson}},\ }\href {\doibase 10.1103/PhysRev.95.228} {\bibfield  {journal}
  {\bibinfo  {journal} {Phys. Rev.}\ }\textbf {\bibinfo {volume} {95}},\
  \bibinfo {pages} {228} (\bibinfo {year} {1954})}\BibitemShut {NoStop}%
\bibitem [{\citenamefont {\L{}ukaszuk}(1973)}]{Lukaszuk:1973jd}%
  \BibitemOpen
  \bibfield  {author} {\bibinfo {author} {\bibfnamefont {L.}~\bibnamefont
  {\L{}ukaszuk}},\ }\href {\doibase 10.1016/0370-2693(73)90567-4} {\bibfield
  {journal} {\bibinfo  {journal} {Phys. Lett.}\ }\textbf {\bibinfo {volume}
  {47B}},\ \bibinfo {pages} {51} (\bibinfo {year} {1973})}\BibitemShut
  {NoStop}%
\bibitem [{\citenamefont {Eidelman}\ and\ \citenamefont
  {\L{}ukaszuk}(2004)}]{Eidelman:2003uh}%
  \BibitemOpen
  \bibfield  {author} {\bibinfo {author} {\bibfnamefont {S.}~\bibnamefont
  {Eidelman}}\ and\ \bibinfo {author} {\bibfnamefont {L.}~\bibnamefont
  {\L{}ukaszuk}},\ }\href {\doibase 10.1016/j.physletb.2003.12.030} {\bibfield
  {journal} {\bibinfo  {journal} {Phys. Lett.}\ }\textbf {\bibinfo {volume}
  {B582}},\ \bibinfo {pages} {27} (\bibinfo {year} {2004})},\ \Eprint
  {http://arxiv.org/abs/hep-ph/0311366} {arXiv:hep-ph/0311366
  [hep-ph]}\BibitemShut {NoStop}%
\bibitem [{\citenamefont {Amendolia}\ \emph {et~al.}(1986)\citenamefont
  {Amendolia} \emph {et~al.}}]{Amendolia:1986wj}%
  \BibitemOpen
  \bibfield  {author} {\bibinfo {author} {\bibfnamefont {S.~R.}\ \bibnamefont
  {Amendolia}} \emph {et~al.} (\bibinfo {collaboration} {NA7}),\ }\href
  {\doibase 10.1016/0550-3213(86)90437-2} {\bibfield  {journal} {\bibinfo
  {journal} {Nucl. Phys.}\ }\textbf {\bibinfo {volume} {B277}},\ \bibinfo
  {pages} {168} (\bibinfo {year} {1986})}\BibitemShut {NoStop}%
\bibitem [{\citenamefont {Akhmetshin}\ \emph {et~al.}(2002)\citenamefont
  {Akhmetshin} \emph {et~al.}}]{Akhmetshin:2001ig}%
  \BibitemOpen
  \bibfield  {author} {\bibinfo {author} {\bibfnamefont {R.~R.}\ \bibnamefont
  {Akhmetshin}} \emph {et~al.} (\bibinfo {collaboration} {CMD-2}),\ }\href
  {\doibase 10.1016/S0370-2693(02)01168-1} {\bibfield  {journal} {\bibinfo
  {journal} {Phys. Lett.}\ }\textbf {\bibinfo {volume} {B527}},\ \bibinfo
  {pages} {161} (\bibinfo {year} {2002})},\ \Eprint
  {http://arxiv.org/abs/hep-ex/0112031} {arXiv:hep-ex/0112031
  [hep-ex]}\BibitemShut {NoStop}%
\bibitem [{\citenamefont {Akhmetshin}\ \emph {et~al.}(2004)\citenamefont
  {Akhmetshin} \emph {et~al.}}]{Akhmetshin:2003zn}%
  \BibitemOpen
  \bibfield  {author} {\bibinfo {author} {\bibfnamefont {R.~R.}\ \bibnamefont
  {Akhmetshin}} \emph {et~al.} (\bibinfo {collaboration} {CMD-2}),\ }\href
  {\doibase 10.1016/j.physletb.2003.10.108} {\bibfield  {journal} {\bibinfo
  {journal} {Phys. Lett.}\ }\textbf {\bibinfo {volume} {B578}},\ \bibinfo
  {pages} {285} (\bibinfo {year} {2004})},\ \Eprint
  {http://arxiv.org/abs/hep-ex/0308008} {arXiv:hep-ex/0308008
  [hep-ex]}\BibitemShut {NoStop}%
\bibitem [{\citenamefont {Achasov}\ \emph {et~al.}(2005)\citenamefont {Achasov}
  \emph {et~al.}}]{Achasov:2005rg}%
  \BibitemOpen
  \bibfield  {author} {\bibinfo {author} {\bibfnamefont {M.~N.}\ \bibnamefont
  {Achasov}} \emph {et~al.},\ }\href {\doibase 10.1134/1.2163921} {\bibfield
  {journal} {\bibinfo  {journal} {J. Exp. Theor. Phys.}\ }\textbf {\bibinfo
  {volume} {101}},\ \bibinfo {pages} {1053} (\bibinfo {year} {2005})},\
  \bibinfo {note} {[Zh. Eksp. Teor. Fiz. {\bf 128}, 1201 (2005)]},\ \Eprint
  {http://arxiv.org/abs/hep-ex/0506076} {arXiv:hep-ex/0506076
  [hep-ex]}\BibitemShut {NoStop}%
\bibitem [{\citenamefont {Achasov}\ \emph {et~al.}(2006)\citenamefont {Achasov}
  \emph {et~al.}}]{Achasov:2006vp}%
  \BibitemOpen
  \bibfield  {author} {\bibinfo {author} {\bibfnamefont {M.~N.}\ \bibnamefont
  {Achasov}} \emph {et~al.},\ }\href {\doibase 10.1134/S106377610609007X}
  {\bibfield  {journal} {\bibinfo  {journal} {J. Exp. Theor. Phys.}\ }\textbf
  {\bibinfo {volume} {103}},\ \bibinfo {pages} {380} (\bibinfo {year}
  {2006})},\ \bibinfo {note} {[Zh. Eksp. Teor. Fiz. {\bf 130}, 437 (2006)]},\
  \Eprint {http://arxiv.org/abs/hep-ex/0605013} {arXiv:hep-ex/0605013
  [hep-ex]}\BibitemShut {NoStop}%
\bibitem [{\citenamefont {Akhmetshin}\ \emph {et~al.}(2006)\citenamefont
  {Akhmetshin} \emph {et~al.}}]{Akhmetshin:2006wh}%
  \BibitemOpen
  \bibfield  {author} {\bibinfo {author} {\bibfnamefont {R.~R.}\ \bibnamefont
  {Akhmetshin}} \emph {et~al.},\ }\href {\doibase 10.1134/S0021364006200021}
  {\bibfield  {journal} {\bibinfo  {journal} {JETP Lett.}\ }\textbf {\bibinfo
  {volume} {84}},\ \bibinfo {pages} {413} (\bibinfo {year} {2006})},\ \bibinfo
  {note} {[Zh. Eksp. Teor. Fiz. {\bf 84}, 491 (2006)]},\ \Eprint
  {http://arxiv.org/abs/hep-ex/0610016} {arXiv:hep-ex/0610016
  [hep-ex]}\BibitemShut {NoStop}%
\bibitem [{\citenamefont {Akhmetshin}\ \emph {et~al.}(2007)\citenamefont
  {Akhmetshin} \emph {et~al.}}]{Akhmetshin:2006bx}%
  \BibitemOpen
  \bibfield  {author} {\bibinfo {author} {\bibfnamefont {R.~R.}\ \bibnamefont
  {Akhmetshin}} \emph {et~al.} (\bibinfo {collaboration} {CMD-2}),\ }\href
  {\doibase 10.1016/j.physletb.2007.01.073} {\bibfield  {journal} {\bibinfo
  {journal} {Phys. Lett.}\ }\textbf {\bibinfo {volume} {B648}},\ \bibinfo
  {pages} {28} (\bibinfo {year} {2007})},\ \Eprint
  {http://arxiv.org/abs/hep-ex/0610021} {arXiv:hep-ex/0610021
  [hep-ex]}\BibitemShut {NoStop}%
\bibitem [{\citenamefont {Ambrosino}\ \emph {et~al.}(2009)\citenamefont
  {Ambrosino} \emph {et~al.}}]{Ambrosino:2008aa}%
  \BibitemOpen
  \bibfield  {author} {\bibinfo {author} {\bibfnamefont {F.}~\bibnamefont
  {Ambrosino}} \emph {et~al.} (\bibinfo {collaboration} {KLOE}),\ }\href
  {\doibase 10.1016/j.physletb.2008.10.060} {\bibfield  {journal} {\bibinfo
  {journal} {Phys. Lett.}\ }\textbf {\bibinfo {volume} {B670}},\ \bibinfo
  {pages} {285} (\bibinfo {year} {2009})},\ \Eprint
  {http://arxiv.org/abs/0809.3950} {arXiv:0809.3950 [hep-ex]}\BibitemShut
  {NoStop}%
\bibitem [{\citenamefont {Aubert}\ \emph {et~al.}(2009)\citenamefont {Aubert}
  \emph {et~al.}}]{Aubert:2009ad}%
  \BibitemOpen
  \bibfield  {author} {\bibinfo {author} {\bibfnamefont {B.}~\bibnamefont
  {Aubert}} \emph {et~al.} (\bibinfo {collaboration} {BaBar}),\ }\href
  {\doibase 10.1103/PhysRevLett.103.231801} {\bibfield  {journal} {\bibinfo
  {journal} {Phys. Rev. Lett.}\ }\textbf {\bibinfo {volume} {103}},\ \bibinfo
  {pages} {231801} (\bibinfo {year} {2009})},\ \Eprint
  {http://arxiv.org/abs/0908.3589} {arXiv:0908.3589 [hep-ex]}\BibitemShut
  {NoStop}%
\bibitem [{\citenamefont {Ambrosino}\ \emph {et~al.}(2011)\citenamefont
  {Ambrosino} \emph {et~al.}}]{Ambrosino:2010bv}%
  \BibitemOpen
  \bibfield  {author} {\bibinfo {author} {\bibfnamefont {F.}~\bibnamefont
  {Ambrosino}} \emph {et~al.} (\bibinfo {collaboration} {KLOE}),\ }\href
  {\doibase 10.1016/j.physletb.2011.04.055} {\bibfield  {journal} {\bibinfo
  {journal} {Phys. Lett.}\ }\textbf {\bibinfo {volume} {B700}},\ \bibinfo
  {pages} {102} (\bibinfo {year} {2011})},\ \Eprint
  {http://arxiv.org/abs/1006.5313} {arXiv:1006.5313 [hep-ex]}\BibitemShut
  {NoStop}%
\bibitem [{\citenamefont {Lees}\ \emph {et~al.}(2012)\citenamefont {Lees} \emph
  {et~al.}}]{Lees:2012cj}%
  \BibitemOpen
  \bibfield  {author} {\bibinfo {author} {\bibfnamefont {J.~P.}\ \bibnamefont
  {Lees}} \emph {et~al.} (\bibinfo {collaboration} {BaBar}),\ }\href {\doibase
  10.1103/PhysRevD.86.032013} {\bibfield  {journal} {\bibinfo  {journal} {Phys.
  Rev.}\ }\textbf {\bibinfo {volume} {D86}},\ \bibinfo {pages} {032013}
  (\bibinfo {year} {2012})},\ \Eprint {http://arxiv.org/abs/1205.2228}
  {arXiv:1205.2228 [hep-ex]}\BibitemShut {NoStop}%
\bibitem [{\citenamefont {Babusci}\ \emph {et~al.}(2013)\citenamefont {Babusci}
  \emph {et~al.}}]{Babusci:2012rp}%
  \BibitemOpen
  \bibfield  {author} {\bibinfo {author} {\bibfnamefont {D.}~\bibnamefont
  {Babusci}} \emph {et~al.} (\bibinfo {collaboration} {KLOE}),\ }\href
  {\doibase 10.1016/j.physletb.2013.02.029} {\bibfield  {journal} {\bibinfo
  {journal} {Phys. Lett.}\ }\textbf {\bibinfo {volume} {B720}},\ \bibinfo
  {pages} {336} (\bibinfo {year} {2013})},\ \Eprint
  {http://arxiv.org/abs/1212.4524} {arXiv:1212.4524 [hep-ex]}\BibitemShut
  {NoStop}%
\bibitem [{\citenamefont {Anastasi}\ \emph {et~al.}(2018)\citenamefont
  {Anastasi} \emph {et~al.}}]{Anastasi:2017eio}%
  \BibitemOpen
  \bibfield  {author} {\bibinfo {author} {\bibfnamefont {A.}~\bibnamefont
  {Anastasi}} \emph {et~al.} (\bibinfo {collaboration} {KLOE-2}),\ }\href
  {\doibase 10.1007/JHEP03(2018)173} {\bibfield  {journal} {\bibinfo  {journal}
  {JHEP}\ }\textbf {\bibinfo {volume} {03}},\ \bibinfo {pages} {173} (\bibinfo
  {year} {2018})},\ \Eprint {http://arxiv.org/abs/1711.03085} {arXiv:1711.03085
  [hep-ex]}\BibitemShut {NoStop}%
\bibitem [{\citenamefont {Feng}\ \emph {et~al.}(2020)\citenamefont {Feng},
  \citenamefont {Fu},\ and\ \citenamefont {Jin}}]{Feng:2019geu}%
  \BibitemOpen
  \bibfield  {author} {\bibinfo {author} {\bibfnamefont {X.}~\bibnamefont
  {Feng}}, \bibinfo {author} {\bibfnamefont {Y.}~\bibnamefont {Fu}}, \ and\
  \bibinfo {author} {\bibfnamefont {L.-C.}\ \bibnamefont {Jin}},\ }\href
  {\doibase 10.1103/PhysRevD.101.051502} {\bibfield  {journal} {\bibinfo
  {journal} {Phys. Rev.}\ }\textbf {\bibinfo {volume} {D101}},\ \bibinfo
  {pages} {051502} (\bibinfo {year} {2020})},\ \Eprint
  {http://arxiv.org/abs/1911.04064} {arXiv:1911.04064 [hep-lat]}\BibitemShut
  {NoStop}%
\bibitem [{\citenamefont {Wang}\ \emph {et~al.}(2020)\citenamefont {Wang},
  \citenamefont {Liang}, \citenamefont {Draper}, \citenamefont {Liu},\ and\
  \citenamefont {Yang}}]{Wang:2020nbf}%
  \BibitemOpen
  \bibfield  {author} {\bibinfo {author} {\bibfnamefont {G.}~\bibnamefont
  {Wang}}, \bibinfo {author} {\bibfnamefont {J.}~\bibnamefont {Liang}},
  \bibinfo {author} {\bibfnamefont {T.}~\bibnamefont {Draper}}, \bibinfo
  {author} {\bibfnamefont {K.-F.}\ \bibnamefont {Liu}}, \ and\ \bibinfo
  {author} {\bibfnamefont {Y.-B.}\ \bibnamefont {Yang}},\ }\href@noop {} {\
  (\bibinfo {year} {2020})},\ \Eprint {http://arxiv.org/abs/2006.05431}
  {arXiv:2006.05431 [hep-ph]}\BibitemShut {NoStop}%
\bibitem [{\citenamefont {de~Rafael}(2020)}]{deRafael:2020uif}%
  \BibitemOpen
  \bibfield  {author} {\bibinfo {author} {\bibfnamefont {E.}~\bibnamefont
  {de~Rafael}},\ }\href {\doibase 10.1103/PhysRevD.102.056025} {\bibfield
  {journal} {\bibinfo  {journal} {Phys. Rev.}\ }\textbf {\bibinfo {volume}
  {D102}},\ \bibinfo {pages} {056025} (\bibinfo {year} {2020})},\ \Eprint
  {http://arxiv.org/abs/2006.13880} {arXiv:2006.13880 [hep-ph]}\BibitemShut
  {NoStop}%
\bibitem [{\citenamefont {Horn}\ \emph {et~al.}(2006)\citenamefont {Horn} \emph
  {et~al.}}]{Horn:2006tm}%
  \BibitemOpen
  \bibfield  {author} {\bibinfo {author} {\bibfnamefont {T.}~\bibnamefont
  {Horn}} \emph {et~al.} (\bibinfo {collaboration} {Jefferson Lab $F_\pi$}),\
  }\href {\doibase 10.1103/PhysRevLett.97.192001} {\bibfield  {journal}
  {\bibinfo  {journal} {Phys. Rev. Lett.}\ }\textbf {\bibinfo {volume} {97}},\
  \bibinfo {pages} {192001} (\bibinfo {year} {2006})},\ \Eprint
  {http://arxiv.org/abs/nucl-ex/0607005} {arXiv:nucl-ex/0607005
  [nucl-ex]}\BibitemShut {NoStop}%
\bibitem [{\citenamefont {Tadevosyan}\ \emph {et~al.}(2007)\citenamefont
  {Tadevosyan} \emph {et~al.}}]{Tadevosyan:2007yd}%
  \BibitemOpen
  \bibfield  {author} {\bibinfo {author} {\bibfnamefont {V.}~\bibnamefont
  {Tadevosyan}} \emph {et~al.} (\bibinfo {collaboration} {Jefferson Lab
  $F_\pi$}),\ }\href {\doibase 10.1103/PhysRevC.75.055205} {\bibfield
  {journal} {\bibinfo  {journal} {Phys. Rev.}\ }\textbf {\bibinfo {volume}
  {C75}},\ \bibinfo {pages} {055205} (\bibinfo {year} {2007})},\ \Eprint
  {http://arxiv.org/abs/nucl-ex/0607007} {arXiv:nucl-ex/0607007
  [nucl-ex]}\BibitemShut {NoStop}%
\bibitem [{\citenamefont {Huber}\ \emph {et~al.}(2008)\citenamefont {Huber}
  \emph {et~al.}}]{Huber:2008id}%
  \BibitemOpen
  \bibfield  {author} {\bibinfo {author} {\bibfnamefont {G.~M.}\ \bibnamefont
  {Huber}} \emph {et~al.} (\bibinfo {collaboration} {Jefferson Lab $F_\pi$}),\
  }\href {\doibase 10.1103/PhysRevC.78.045203} {\bibfield  {journal} {\bibinfo
  {journal} {Phys. Rev.}\ }\textbf {\bibinfo {volume} {C78}},\ \bibinfo {pages}
  {045203} (\bibinfo {year} {2008})},\ \Eprint {http://arxiv.org/abs/0809.3052}
  {arXiv:0809.3052 [nucl-ex]}\BibitemShut {NoStop}%
\bibitem [{\citenamefont {Blok}\ \emph {et~al.}(2008)\citenamefont {Blok} \emph
  {et~al.}}]{Blok:2008jy}%
  \BibitemOpen
  \bibfield  {author} {\bibinfo {author} {\bibfnamefont {H.~P.}\ \bibnamefont
  {Blok}} \emph {et~al.} (\bibinfo {collaboration} {Jefferson Lab $F_\pi$}),\
  }\href {\doibase 10.1103/PhysRevC.78.045202} {\bibfield  {journal} {\bibinfo
  {journal} {Phys. Rev.}\ }\textbf {\bibinfo {volume} {C78}},\ \bibinfo {pages}
  {045202} (\bibinfo {year} {2008})},\ \Eprint {http://arxiv.org/abs/0809.3161}
  {arXiv:0809.3161 [nucl-ex]}\BibitemShut {NoStop}%
\bibitem [{\citenamefont {Wilson}\ \emph {et~al.}(2015)\citenamefont {Wilson},
  \citenamefont {Brice{\~n}o}, \citenamefont {Dudek}, \citenamefont {Edwards},\
  and\ \citenamefont {Thomas}}]{Wilson:2015dqa}%
  \BibitemOpen
  \bibfield  {author} {\bibinfo {author} {\bibfnamefont {D.~J.}\ \bibnamefont
  {Wilson}}, \bibinfo {author} {\bibfnamefont {R.~A.}\ \bibnamefont
  {Brice{\~n}o}}, \bibinfo {author} {\bibfnamefont {J.~J.}\ \bibnamefont
  {Dudek}}, \bibinfo {author} {\bibfnamefont {R.~G.}\ \bibnamefont {Edwards}},
  \ and\ \bibinfo {author} {\bibfnamefont {C.~E.}\ \bibnamefont {Thomas}},\
  }\href {\doibase 10.1103/PhysRevD.92.094502} {\bibfield  {journal} {\bibinfo
  {journal} {Phys. Rev.}\ }\textbf {\bibinfo {volume} {D92}},\ \bibinfo {pages}
  {094502} (\bibinfo {year} {2015})},\ \Eprint
  {http://arxiv.org/abs/1507.02599} {arXiv:1507.02599 [hep-ph]}\BibitemShut
  {NoStop}%
\bibitem [{\citenamefont {Bali}\ \emph {et~al.}(2016)\citenamefont {Bali},
  \citenamefont {Collins}, \citenamefont {Cox}, \citenamefont {Donald},
  \citenamefont {G\"ockeler}, \citenamefont {Lang},\ and\ \citenamefont
  {Sch\"afer}}]{Bali:2015gji}%
  \BibitemOpen
  \bibfield  {author} {\bibinfo {author} {\bibfnamefont {G.~S.}\ \bibnamefont
  {Bali}}, \bibinfo {author} {\bibfnamefont {S.}~\bibnamefont {Collins}},
  \bibinfo {author} {\bibfnamefont {A.}~\bibnamefont {Cox}}, \bibinfo {author}
  {\bibfnamefont {G.}~\bibnamefont {Donald}}, \bibinfo {author} {\bibfnamefont
  {M.}~\bibnamefont {G\"ockeler}}, \bibinfo {author} {\bibfnamefont
  {C.}~\bibnamefont {Lang}}, \ and\ \bibinfo {author} {\bibfnamefont
  {A.}~\bibnamefont {Sch\"afer}} (\bibinfo {collaboration} {RQCD}),\ }\href
  {\doibase 10.1103/PhysRevD.93.054509} {\bibfield  {journal} {\bibinfo
  {journal} {Phys. Rev.}\ }\textbf {\bibinfo {volume} {D93}},\ \bibinfo {pages}
  {054509} (\bibinfo {year} {2016})},\ \Eprint
  {http://arxiv.org/abs/1512.08678} {arXiv:1512.08678 [hep-lat]}\BibitemShut
  {NoStop}%
\bibitem [{\citenamefont {Guo}\ \emph {et~al.}(2016)\citenamefont {Guo},
  \citenamefont {Alexandru}, \citenamefont {Molina},\ and\ \citenamefont
  {D\"oring}}]{Guo:2016zos}%
  \BibitemOpen
  \bibfield  {author} {\bibinfo {author} {\bibfnamefont {D.}~\bibnamefont
  {Guo}}, \bibinfo {author} {\bibfnamefont {A.}~\bibnamefont {Alexandru}},
  \bibinfo {author} {\bibfnamefont {R.}~\bibnamefont {Molina}}, \ and\ \bibinfo
  {author} {\bibfnamefont {M.}~\bibnamefont {D\"oring}},\ }\href {\doibase
  10.1103/PhysRevD.94.034501} {\bibfield  {journal} {\bibinfo  {journal} {Phys.
  Rev.}\ }\textbf {\bibinfo {volume} {D94}},\ \bibinfo {pages} {034501}
  (\bibinfo {year} {2016})},\ \Eprint {http://arxiv.org/abs/1605.03993}
  {arXiv:1605.03993 [hep-lat]}\BibitemShut {NoStop}%
\bibitem [{\citenamefont {Fu}\ and\ \citenamefont {Wang}(2016)}]{Fu:2016itp}%
  \BibitemOpen
  \bibfield  {author} {\bibinfo {author} {\bibfnamefont {Z.}~\bibnamefont
  {Fu}}\ and\ \bibinfo {author} {\bibfnamefont {L.}~\bibnamefont {Wang}},\
  }\href {\doibase 10.1103/PhysRevD.94.034505} {\bibfield  {journal} {\bibinfo
  {journal} {Phys. Rev.}\ }\textbf {\bibinfo {volume} {D94}},\ \bibinfo {pages}
  {034505} (\bibinfo {year} {2016})},\ \Eprint
  {http://arxiv.org/abs/1608.07478} {arXiv:1608.07478 [hep-lat]}\BibitemShut
  {NoStop}%
\bibitem [{\citenamefont {Alexandrou}\ \emph {et~al.}(2017)\citenamefont
  {Alexandrou}, \citenamefont {Leskovec}, \citenamefont {Meinel}, \citenamefont
  {Negele}, \citenamefont {Paul}, \citenamefont {Petschlies}, \citenamefont
  {Pochinsky}, \citenamefont {Rendon},\ and\ \citenamefont
  {Syritsyn}}]{Alexandrou:2017mpi}%
  \BibitemOpen
  \bibfield  {author} {\bibinfo {author} {\bibfnamefont {C.}~\bibnamefont
  {Alexandrou}}, \bibinfo {author} {\bibfnamefont {L.}~\bibnamefont
  {Leskovec}}, \bibinfo {author} {\bibfnamefont {S.}~\bibnamefont {Meinel}},
  \bibinfo {author} {\bibfnamefont {J.}~\bibnamefont {Negele}}, \bibinfo
  {author} {\bibfnamefont {S.}~\bibnamefont {Paul}}, \bibinfo {author}
  {\bibfnamefont {M.}~\bibnamefont {Petschlies}}, \bibinfo {author}
  {\bibfnamefont {A.}~\bibnamefont {Pochinsky}}, \bibinfo {author}
  {\bibfnamefont {G.}~\bibnamefont {Rendon}}, \ and\ \bibinfo {author}
  {\bibfnamefont {S.}~\bibnamefont {Syritsyn}},\ }\href {\doibase
  10.1103/PhysRevD.96.034525} {\bibfield  {journal} {\bibinfo  {journal} {Phys.
  Rev.}\ }\textbf {\bibinfo {volume} {D96}},\ \bibinfo {pages} {034525}
  (\bibinfo {year} {2017})},\ \Eprint {http://arxiv.org/abs/1704.05439}
  {arXiv:1704.05439 [hep-lat]}\BibitemShut {NoStop}%
\bibitem [{\citenamefont {Andersen}\ \emph {et~al.}(2019)\citenamefont
  {Andersen}, \citenamefont {Bulava}, \citenamefont {H{\"o}rz},\ and\
  \citenamefont {Morningstar}}]{Andersen:2018mau}%
  \BibitemOpen
  \bibfield  {author} {\bibinfo {author} {\bibfnamefont {C.}~\bibnamefont
  {Andersen}}, \bibinfo {author} {\bibfnamefont {J.}~\bibnamefont {Bulava}},
  \bibinfo {author} {\bibfnamefont {B.}~\bibnamefont {H{\"o}rz}}, \ and\
  \bibinfo {author} {\bibfnamefont {C.}~\bibnamefont {Morningstar}},\ }\href
  {\doibase 10.1016/j.nuclphysb.2018.12.018} {\bibfield  {journal} {\bibinfo
  {journal} {Nucl. Phys.}\ }\textbf {\bibinfo {volume} {B939}},\ \bibinfo
  {pages} {145} (\bibinfo {year} {2019})},\ \Eprint
  {http://arxiv.org/abs/1808.05007} {arXiv:1808.05007 [hep-lat]}\BibitemShut
  {NoStop}%
\bibitem [{\citenamefont {Werner}\ \emph {et~al.}(2020)\citenamefont {Werner}
  \emph {et~al.}}]{Werner:2019hxc}%
  \BibitemOpen
  \bibfield  {author} {\bibinfo {author} {\bibfnamefont {M.}~\bibnamefont
  {Werner}} \emph {et~al.} (\bibinfo {collaboration} {Extended Twisted Mass}),\
  }\href {\doibase 10.1140/epja/s10050-020-00057-4} {\bibfield  {journal}
  {\bibinfo  {journal} {Eur. Phys. J.}\ }\textbf {\bibinfo {volume} {A56}},\
  \bibinfo {pages} {61} (\bibinfo {year} {2020})},\ \Eprint
  {http://arxiv.org/abs/1907.01237} {arXiv:1907.01237 [hep-lat]}\BibitemShut
  {NoStop}%
\bibitem [{\citenamefont {Erben}\ \emph {et~al.}(2020)\citenamefont {Erben},
  \citenamefont {Green}, \citenamefont {Mohler},\ and\ \citenamefont
  {Wittig}}]{Erben:2019nmx}%
  \BibitemOpen
  \bibfield  {author} {\bibinfo {author} {\bibfnamefont {F.}~\bibnamefont
  {Erben}}, \bibinfo {author} {\bibfnamefont {J.~R.}\ \bibnamefont {Green}},
  \bibinfo {author} {\bibfnamefont {D.}~\bibnamefont {Mohler}}, \ and\ \bibinfo
  {author} {\bibfnamefont {H.}~\bibnamefont {Wittig}},\ }\href {\doibase
  10.1103/PhysRevD.101.054504} {\bibfield  {journal} {\bibinfo  {journal}
  {Phys. Rev.}\ }\textbf {\bibinfo {volume} {D101}},\ \bibinfo {pages} {054504}
  (\bibinfo {year} {2020})},\ \Eprint {http://arxiv.org/abs/1910.01083}
  {arXiv:1910.01083 [hep-lat]}\BibitemShut {NoStop}%
\bibitem [{\citenamefont {Fischer}\ \emph {et~al.}(2020)\citenamefont
  {Fischer}, \citenamefont {Kostrzewa}, \citenamefont {Mai}, \citenamefont
  {Petschlies}, \citenamefont {Pittler}, \citenamefont {Ueding}, \citenamefont
  {Urbach},\ and\ \citenamefont {Werner}}]{Fischer:2020fvl}%
  \BibitemOpen
  \bibfield  {author} {\bibinfo {author} {\bibfnamefont {M.}~\bibnamefont
  {Fischer}}, \bibinfo {author} {\bibfnamefont {B.}~\bibnamefont {Kostrzewa}},
  \bibinfo {author} {\bibfnamefont {M.}~\bibnamefont {Mai}}, \bibinfo {author}
  {\bibfnamefont {M.}~\bibnamefont {Petschlies}}, \bibinfo {author}
  {\bibfnamefont {F.}~\bibnamefont {Pittler}}, \bibinfo {author} {\bibfnamefont
  {M.}~\bibnamefont {Ueding}}, \bibinfo {author} {\bibfnamefont
  {C.}~\bibnamefont {Urbach}}, \ and\ \bibinfo {author} {\bibfnamefont
  {M.}~\bibnamefont {Werner}} (\bibinfo {collaboration} {ETM}),\ }\href@noop {}
  {\  (\bibinfo {year} {2020})},\ \Eprint {http://arxiv.org/abs/2006.13805}
  {arXiv:2006.13805 [hep-lat]}\BibitemShut {NoStop}%
\bibitem [{\citenamefont {Niehus}\ \emph {et~al.}(2020)\citenamefont {Niehus},
  \citenamefont {Hoferichter}, \citenamefont {Kubis},\ and\ \citenamefont
  {Ruiz~de Elvira}}]{Niehus:2020gmf}%
  \BibitemOpen
  \bibfield  {author} {\bibinfo {author} {\bibfnamefont {M.}~\bibnamefont
  {Niehus}}, \bibinfo {author} {\bibfnamefont {M.}~\bibnamefont {Hoferichter}},
  \bibinfo {author} {\bibfnamefont {B.}~\bibnamefont {Kubis}}, \ and\ \bibinfo
  {author} {\bibfnamefont {J.}~\bibnamefont {Ruiz~de Elvira}},\ }\href@noop {}
  {\  (\bibinfo {year} {2020})},\ \Eprint {http://arxiv.org/abs/2009.04479}
  {arXiv:2009.04479 [hep-ph]}\BibitemShut {NoStop}%
\bibitem [{\citenamefont {Aul'chenko}\ \emph {et~al.}(2005)\citenamefont
  {Aul'chenko} \emph {et~al.}}]{Aulchenko:2006na}%
  \BibitemOpen
  \bibfield  {author} {\bibinfo {author} {\bibfnamefont {V.}~\bibnamefont
  {Aul'chenko}} \emph {et~al.} (\bibinfo {collaboration} {CMD-2}),\ }\href
  {\doibase 10.1134/1.2175241} {\bibfield  {journal} {\bibinfo  {journal} {JETP
  Lett.}\ }\textbf {\bibinfo {volume} {82}},\ \bibinfo {pages} {743} (\bibinfo
  {year} {2005})},\ \Eprint {http://arxiv.org/abs/hep-ex/0603021}
  {arXiv:hep-ex/0603021}\BibitemShut {NoStop}%
\bibitem [{\citenamefont {Fujikawa}\ \emph {et~al.}(2008)\citenamefont
  {Fujikawa} \emph {et~al.}}]{Fujikawa:2008ma}%
  \BibitemOpen
  \bibfield  {author} {\bibinfo {author} {\bibfnamefont {M.}~\bibnamefont
  {Fujikawa}} \emph {et~al.} (\bibinfo {collaboration} {Belle}),\ }\href
  {\doibase 10.1103/PhysRevD.78.072006} {\bibfield  {journal} {\bibinfo
  {journal} {Phys. Rev.}\ }\textbf {\bibinfo {volume} {D78}},\ \bibinfo {pages}
  {072006} (\bibinfo {year} {2008})},\ \Eprint {http://arxiv.org/abs/0805.3773}
  {arXiv:0805.3773 [hep-ex]}\BibitemShut {NoStop}%
\bibitem [{\citenamefont {Ablikim}\ \emph {et~al.}(2016)\citenamefont {Ablikim}
  \emph {et~al.}}]{Ablikim:2015orh}%
  \BibitemOpen
  \bibfield  {author} {\bibinfo {author} {\bibfnamefont {M.}~\bibnamefont
  {Ablikim}} \emph {et~al.} (\bibinfo {collaboration} {BESIII}),\ }\href
  {\doibase 10.1016/j.physletb.2015.11.043} {\bibfield  {journal} {\bibinfo
  {journal} {Phys. Lett.}\ }\textbf {\bibinfo {volume} {B753}},\ \bibinfo
  {pages} {629} (\bibinfo {year} {2016})},\ \Eprint
  {http://arxiv.org/abs/1507.08188} {arXiv:1507.08188 [hep-ex]}\BibitemShut
  {NoStop}%
\bibitem [{\citenamefont {Ablikim}\ \emph {et~al.}(2020)\citenamefont {Ablikim}
  \emph {et~al.}}]{Ablikim:2020bah}%
  \BibitemOpen
  \bibfield  {author} {\bibinfo {author} {\bibfnamefont {M.}~\bibnamefont
  {Ablikim}} \emph {et~al.} (\bibinfo {collaboration} {BESIII}),\ }\href@noop
  {} {\  (\bibinfo {year} {2020})},\ \Eprint {http://arxiv.org/abs/2009.05011}
  {arXiv:2009.05011 [hep-ex]}\BibitemShut {NoStop}%
\bibitem [{\citenamefont {Achasov}\ \emph {et~al.}(2020)\citenamefont {Achasov}
  \emph {et~al.}}]{Achasov:2020iys}%
  \BibitemOpen
  \bibfield  {author} {\bibinfo {author} {\bibfnamefont {M.}~\bibnamefont
  {Achasov}} \emph {et~al.} (\bibinfo {collaboration} {SND}),\ }\href@noop {}
  {\  (\bibinfo {year} {2020})},\ \Eprint {http://arxiv.org/abs/2004.00263}
  {arXiv:2004.00263 [hep-ex]}\BibitemShut {NoStop}%
\bibitem [{\citenamefont {Ignatov}\ \emph {et~al.}(2019)\citenamefont {Ignatov}
  \emph {et~al.}}]{Ignatov:2019omb}%
  \BibitemOpen
  \bibfield  {author} {\bibinfo {author} {\bibfnamefont {F.}~\bibnamefont
  {Ignatov}} \emph {et~al.} (\bibinfo {collaboration} {{CMD-3}}),\ }\href
  {\doibase 10.1051/epjconf/201921204001} {\bibfield  {journal} {\bibinfo
  {journal} {EPJ Web Conf.}\ }\textbf {\bibinfo {volume} {212}},\ \bibinfo
  {pages} {04001} (\bibinfo {year} {2019})}\BibitemShut {NoStop}%
\bibitem [{\citenamefont {Abbiendi}\ \emph {et~al.}(2017)\citenamefont
  {Abbiendi} \emph {et~al.}}]{Abbiendi:2016xup}%
  \BibitemOpen
  \bibfield  {author} {\bibinfo {author} {\bibfnamefont {G.}~\bibnamefont
  {Abbiendi}} \emph {et~al.},\ }\href {\doibase 10.1140/epjc/s10052-017-4633-z}
  {\bibfield  {journal} {\bibinfo  {journal} {Eur. Phys. J.}\ }\textbf
  {\bibinfo {volume} {C77}},\ \bibinfo {pages} {139} (\bibinfo {year}
  {2017})},\ \Eprint {http://arxiv.org/abs/1609.08987} {arXiv:1609.08987
  [hep-ex]}\BibitemShut {NoStop}%
\bibitem [{\citenamefont {Banerjee}\ \emph {et~al.}(2020)\citenamefont
  {Banerjee} \emph {et~al.}}]{Banerjee:2020tdt}%
  \BibitemOpen
  \bibfield  {author} {\bibinfo {author} {\bibfnamefont {P.}~\bibnamefont
  {Banerjee}} \emph {et~al.},\ }\href {\doibase 10.1140/epjc/s10052-020-8138-9}
  {\bibfield  {journal} {\bibinfo  {journal} {Eur. Phys. J.}\ }\textbf
  {\bibinfo {volume} {C80}},\ \bibinfo {pages} {591} (\bibinfo {year}
  {2020})},\ \Eprint {http://arxiv.org/abs/2004.13663} {arXiv:2004.13663
  [hep-ph]}\BibitemShut {NoStop}%
\end{thebibliography}%

\end{document}